\documentclass[%
 reprint,
 amsmath,amssymb,
 aps,
 prd
]{revtex4-1}

\usepackage{graphicx}
\usepackage{dcolumn}
\usepackage{bm}
\usepackage{float}
\usepackage{natbib}
\usepackage{multirow}
\usepackage[caption=false]{subfig}
\usepackage{lineno}
\usepackage[english, status=draft]{fixme}
\fxusetheme{color}
\usepackage{amsmath}
\usepackage{needspace}
\usepackage{array}
\usepackage{booktabs}
\usepackage{subdepth}
\usepackage{algorithm}
\usepackage[noend]{algpseudocode}
\usepackage{lineno}
\usepackage[normalem]{ulem}
\usepackage[utf8]{inputenc}
\usepackage[all]{nowidow}

\usepackage{color}

\newcolumntype{M}[1]{>{\centering\arraybackslash}m{#1}}
\newcommand{\GP}{{\small GP}}
\newcommand{\GPR}{{\small GPR}}
\newcommand{\GW}{{\small GW}}

\newcommand{\NR}{{\small NR}}
\newcommand{\PE}{{\small PE}}

\newcommand{\CBC}{{\small CBC}}
\newcommand{\params}{\vec{\lambda}}
\newcommand{\ROM}{{\small ROM}}
\newcommand{\beq}{\begin{equation}}
\newcommand{\eeq}{\end{equation}}
\newcommand{\bbm}{\begin{bmatrix}}
\newcommand{\ebm}{\end{bmatrix}}
\newcommand{\sn}{\times10^}
\newcommand{\ntrain}{n_{\rm train}}
\newcommand{\ninterp}{n_{\rm interp}}

\newcommand{\LIGO}{{\small LIGO}}

\begin{document}

\preprint{APS/123-QED}

\title{Statistical Gravitational Waveform Models: What to Simulate Next?}

\author{Zoheyr Doctor}
\affiliation{Kavli Institute for Cosmological Physics, University of Chicago, Chicago, IL 60637, USA}
\affiliation{Department of Physics, University of Chicago, Chicago, IL 60637, USA}

\author{Ben Farr}
\affiliation{Kavli Institute for Cosmological Physics, University of Chicago, Chicago, IL 60637, USA}
\affiliation{Department of Physics, University of Chicago, Chicago, IL 60637, USA}
\affiliation{Enrico Fermi Institute, University of Chicago, Chicago, IL 60637, USA}
\author{Daniel E. Holz}
\affiliation{Kavli Institute for Cosmological Physics, University of Chicago, Chicago, IL 60637, USA}
\affiliation{Department of Physics, University of Chicago, Chicago, IL 60637, USA}
\affiliation{Enrico Fermi Institute, University of Chicago, Chicago, IL 60637, USA}
\affiliation{Department of Astronomy and Astrophysics, University of Chicago, Chicago, IL 60637, USA}
\author{Michael P\"urrer}
\affiliation{Max Planck Institute for Gravitational Physics (Albert Einstein Institute), Am M\"uhlenberg 1, Potsdam 14476, Germany}



\date{\today}

\begin{abstract}
Models of gravitational waveforms play a critical role in detecting and characterizing the gravitational waves (\GW s) from compact binary coalescences.  Waveforms from numerical relativity (\NR), while highly accurate, are too computationally expensive to produce to be directly used with Bayesian parameter estimation tools like Markov-chain-Monte-Carlo and nested sampling. We propose a Gaussian process regression (\GPR) method to generate accurate reduced-order-model waveforms based only on existing accurate (e.g. \NR{}) simulations. Using a training set of simulated waveforms, our \GPR{} approach produces interpolated waveforms along with uncertainties across the parameter space. As a proof of concept, we use a training set of IMRPhenomD waveforms to build a \GPR{} model in the 2-d parameter space of mass ratio $q$ and equal-and-aligned spin $\chi_1=\chi_2$. Using a regular, equally-spaced grid of 120 IMRPhenomD training waveforms in $q\in[1,3]$ and $\chi_1 \in [-0.5,0.5]$, the \GPR{} mean approximates IMRPhenomD in this space to mismatches below $4.3\sn{-5}$. Our approach can alternatively use training waveforms directly from numerical relativity. Beyond interpolation of waveforms, we also present a greedy algorithm that utilizes the errors provided by our \GPR{} model to optimize the placement of future simulations. In a fiducial test case we find that using the greedy algorithm to iteratively add simulations achieves \GPR{} errors that are $\sim 1$ order of magnitude lower than the errors from using Latin-hypercube or square training grids.
\end{abstract}

\pacs{Valid PACS appear here}
\maketitle

\section{Introduction}
\label{sec:intro}
The advent of gravitational wave (\GW) detections has reinforced the need for accurate \GW{} waveform models.  The discovery and parameter estimates of the first LIGO detections were based on matched-filtering techniques which compared the data with predicted \GW{} waveforms \citep{LIGO150914,GW151226,GW170104}.  Parameter estimation (\PE) of these events has already revealed key astrophysical insights \citep{PE150914,astro150914}.  These inferences depend crucially on robust signal detection and parameter estimation, which in turn depend on accurate signal modeling.

Ideally, \PE{} studies with matched-filtering analyses would directly use solutions to Einstein's equations for comparison to observed strain data.  At present the most accurate solutions to Einstein's equations for compact binary coalescences (\CBC{s}) come from numerical relativity (\NR).  Some of these \NR{} simulations have been compared directly to GW150914 and GW151226, with excellent agreement \citep{Lovelace,GW151226, Abbott:2016apu}.  While numerical relativity solutions are accurate, they can take weeks to months to compute \citep{Lovelace}, and thus are prohibitively computationally expensive for use in \PE. For binary black holes, \PE{} requires waveform models to cover a 7-d parameter space. For circularized binary black holes, there are 15 source parameters which need to be estimated, eight of which are intrinsic to the binary (each black hole has a 3-d spin vector and a mass). A shift in the binary total mass simply results in a shift in the gravitational waveform frequency, so in practice the strain need only be solved as a function of the black-hole spins and the binary's mass ratio. Since each \NR{} simulation can take many weeks to run, densely filling the 7-d mass-ratio + spin parameter space with \NR{} waveforms for matched-filter analyses is inconceivable with current resources.

To circumvent the computational expense of \NR{} simulations, approximate waveforms (``approximants", e.g. \citep{Husa:2015iqa, Khan:2015jqa, Taracchini:2013rva, Babak:2016tgq,Hannam,Bohe:2017}) are instead used for \PE{} \citep{PE150914}.  These approximants vary in accuracy, computational expense, source parameter domain, and approximation method, which can limit the potential breadth of \PE{} searches or result in biased parameter estimates~\citep{Abbott:2016wiq}. The SEOBNRv3 approximant, for example, is the most complete in that it models a 7-d intrinsic parameter space for \CBC{}s in the effective one-body framework, but it has only been calibrated to a few non-precessing \NR{} simulations and requires significant computational resources to generate waveforms \citep{Pan}. Data-driven models exist as well, such as \NR{} surrogates, which fit to existing \NR{} simulations to facilitate interpolation of waveforms at new parameter points \citep{Blackman,BlackmanPrecess}.  As computational resources for generating \NR{} waveforms increase, data-driven models can more readily be used for \PE, in addition to calibration and testing of approximate models (e.g. \citep{oshaughnessy}).


Here we propose a method to obtain reduced-order-model (\ROM) waveforms and ``interpolation'' uncertainties using only a training set of simulations at a small number of points in parameter space.  The impetus for using \ROM{}s, which are concise representations of waveforms, is that our method interpolates between waveforms ``observed" in simulations and hence benefits computationally from reduced waveform dimensionality.  That is, if a time-series waveform with $N$ points can instead be represented by a list of $m<N$ features, only $m$, rather than $N$, evaluations are required to produce a gravitational waveform with source parameters $\vec{\lambda}$.  Here we employ an interpolation technique known as Gaussian process regression (\GPR). The advantage of a Gaussian process (\GP{}) method is that it is {\it statistical} and fast. We describe our basic method in \S\ref{sec:method}, which largely replicates the \ROM{} work of \citep{Purrer2014}, but instead of using spline interpolation to provide point predictions we use \GPR{} to provide a statistical interpolation with uncertainties.  The primary application of our method would be to use \NR{} simulations to train a flexible, \NR-driven \GPR{} waveform model with quantified uncertainties.  Although the end goal is to use \NR{} simulations for training, in this paper we only present a proof of concept. Rather than \NR{} waveforms, we use approximant waveforms from IMRPhenomD \citep{Khan} to train and cross-validate, since they can be generated quickly and for a wide range of parameter values. \S\ref{sec:results} details the results of cross-validating \GPR{} models trained on a small number of IMRPhenomD waveforms with the IMRPhenomD waveforms themselves. We find that the \GPR{} model, trained on just a small subsample of IMRPhenomD waveforms, is able to reproduce IMRPhenomD waveforms to excellent accuracy.  Future work will implement an \NR{} training set in the \GPR{} model. 

One major advantage of the use of \GPR{} methods is that they naturally provide estimates of the errors in the resulting waveforms. Current waveform approximants used in \LIGO{} \PE{} are implicitly assumed to be perfect, although multiple approximants are used to assess systematic errors. A more refined analysis, leading to improved \PE{} results, would incorporate errors in the waveform approximants, especially as a function of location in parameter space. The \GPR{} methods naturally provide a statistically consistent estimate of these errors.  \citep{MooreGair} and \citep{MooreBerry} proposed an improvement to \PE{} in which \GPR{} is used to infer the systematic error on an approximant over the parameter space. They train their \GP{} on the systematic errors between approximants and simulations for parameters where simulations exist. The systematic errors inferred from \GPR{} are then propagated to the likelihood function used in \PE.   A proof of concept of this systematic error interpolation method has only been done in 1-d, and it requires an existing approximant off of which to interpolate the errors.  In contrast, our method requires no other approximants.  Additionally, we consider both 1- and $\mbox{2-d}$ in this work, though extending to more dimensions is straightforward in principle. 

A further advantage of GPR methods and the associated waveform error estimates is that these can be used to optimize the placement of new simulations which can be added to the training set. In \S\ref{sec:optimize} we describe a method for estimating where in parameter space the \GPR{} model has high error, based only on the \GPR{} uncertainties.  This estimate is a natural metric for ``greedily" deciding where new simulations are needed to minimize \GPR{} model errors. In effect, our method will tell you the ``optimal" place in parameter space to run the next simulation. An iterative procedure of \GPR{} building and \NR{} simulation leads to an efficient training set and \GPR{} interpolation based on fewer training points than a naive regular grid. We outline that iterative procedure here:
\begin{enumerate}
\item Use existing (\NR) simulations to build a \GPR{} model with uncertainties.
\item Use \GPR{} uncertainties to estimate where in parameter space errors in the \GPR{} model are highest.
\item Generate new simulations at parameter values with high estimated \GPR{} error and rebuild a \GPR{} model with the augmented training set that includes the new simulations.
\item Repeat.
\end{enumerate}

In \S\ref{sec:discussion} we discuss alternate choices that could be made with respect to \GPR{} modeling of \GW{}s, as the ideas presented herein serve more as a framework for \GPR{} \GW{} models than as an immutable, specific method.  We also consider the evaluation time of \GPR{} models, which vary in speed depending on the number of training waveforms and \ROM{} coefficients.  We conclude in \S\ref{sec:conclusion}.   

\section{Method}
\label{sec:method}
\subsection{Outline for building GPR models}
To minimize the computational expense of waveform interpolation at source parameters $\vec{\lambda}$, we represent the waveforms with a \ROM.  We utilize the method described in \citet{Purrer2014} (hereafter P14), but other choices for building \ROM{}s could also be made (e.g.~\citep{Field}). The steps we take to create a \GPR{} waveform interpolation model based on simulations are:
\begin{enumerate}
\item Simulate $n_{\rm train}$ frequency-domain waveforms at $\params = \{\params_j\}_{j=1}^{n_{\rm train}}$ to cover the parameter space of interest for use as the training set. 

\item Project waveforms into a reduced-order basis so that a waveform with parameters $\params$ is described by a list of coefficients $\{c_i(\params)\}_{i=1}^m$, where $m = {\rm{dim(basis)}}$.

\item \label{item:regularize} Regularize each $c_i$ function by subtracting a linear fit of the training values $c_i(\{\params_j\})$ and normalizing. Define $\tilde{c}_i \equiv {\rm Regularize}(c_i)$. 

\item Assume each $\tilde{c}_i(\params)$ is a realization of a Gaussian process, and use $\tilde{c}_{i}(\{\params_j\})$ as the training set to regress $\tilde{c}_{i}(\params) \approx \tilde{c}_{i,{\rm GP}}(\params)$.

\item Transform $\{\tilde{c}_{i}(\params)\}$ and their uncertainties to the domain of interest (e.g. frequency domain for \LIGO{} \PE{}) by applying the reverse of the regularizations in step \ref{item:regularize} and then projecting the coefficients out of the \ROM{} basis back to the time or frequency domain.  The specific operations to go from the \ROM{} to the time or frequency domain depend on which \ROM{} is used.
\end{enumerate}
We discuss all of these steps in detail below.

\subsection{Waveform Generation and Representation}
\label{subsec:WFgen}
We generate fixed-chirp-mass, frequency-domain IMRPhenomD waveforms (with a starting frequency of 20 Hz) at various source parameter values, and interpolate the amplitudes and phases separately onto sparse frequency grids as in \S5.1 of P14\footnote{We interpolate fixed-chirp-mass waveforms, whereas P14 interpolated fixed-total-mass waveforms.  This results in interpolations which are over the same frequency range rather than interpolations over the same geometric frequency range.}.  For simplicity, only the $h_+$ components of the waveforms are considered here, but the methods presented herein can be identically applied to $h_\times$, or $h_+$ can be used to calculate $h_\times$ as in equations~6.14 and 6.15 in P14. For the sparse frequency grid, we choose an amplitude grid spacing at frequency $f$ of $\Delta_A = 0.1f$ and a phase grid spacing of $\Delta_\Phi = 0.3f^{4/3}$; P14 found that these grid spacings keep a constant spline interpolation error at all frequencies.  The interpolated amplitudes and phases are then packed into the columns of matrices $\mathcal{T}_A$, $\mathcal{T}_\Phi$, and a SVD is performed on each matrix (see \S6 of P14):
\beq
\begin{aligned}
\mathcal{T}_A &= V_A \Sigma_A U^\top_A, \\
\mathcal{T}_\Phi &= V_\Phi \Sigma_\Phi U^\top_\Phi.
\end{aligned}
\eeq
P14 truncates the $V$ matrices to reduce the dimensionality of the \ROM, but for simplicity we instead directly compute projection coefficients $c(\tau)$ for each input amplitude or phase $\tau$ (amplitude and phase subscripts dropped for ease of notation):
\beq
c(\tau) = V^\top\tau.
\eeq
Each element of $c$ is a function of $\tau$ and hence a function of the input parameters $\params$.  The following sections describe how the elements of $c$ are interpolated across the parameter space to extract new waveforms with uncertainties.

\subsection{Gaussian Process Regression}
\GPR{} is a tool for emulating (i.e., statistically inferring) the behavior of functions of continuous variables.  It is commonly used to predict the output of simulations which are too expensive to run for many parameter values.  In the case of \GW{s}, we wish to infer new strain waveforms from existing \NR{} simulations. The basic assumption in \GPR{} is that any finite subset of values of a process $f(\vec{x})$ have a joint Gaussian distribution and thus can be described by a mean function $\vec{\mu}$ and covariance function $\mathbf{k}(\vec{x},\vec{x} ')$:
\beq
\label{eqn:GPPrior}
f(\vec{x}) \sim \mathcal{GP}(\vec{\mu},\mathbf{k}(\vec{x},\vec{x} ')).
\eeq 
Given a list of known values of a function $\mathbf{f}=\{f_1(\vec{x}_1), f_2(\vec{x}_2),...\}$ at $n_{\rm train}$ training points $X=\{\vec{x}_1, \vec{x}_2,...\}$, we can calculate the probability distribution for $\mathbf{f}_* = \{f_{1*}(\vec{x}_{1*}), f_{2*}(\vec{x}_{2*}), ...\}$ at new points $X_*=\{\vec{x}_{1*}, \vec{x}_{2*},...\}$, using the definition of a Gaussian process: the prediction and the known values have a joint Gaussian distribution.  If the mean of the \GP{} prior in Equation \ref{eqn:GPPrior} is zero\footnote{We assume the prior mean is zero, as we only interpolate functions which are de-meaned, i.e. which have their mean subtracted off.} then:
\beq
\bbm \mathbf{f} \\ \mathbf{f}_* \ebm = \mathcal{N}\left(\mathbf{0}, \bbm K(X,X) & K(X,X_*) \\ K(X_*,X) & K(X_*,X_*) \ebm \right),
\eeq
where $\mathcal{N}(\mathbf{m}, \mathbf{K})$ is a multivariate normal distribution with mean $\mathbf{m}$ and covariance matrix $\mathbf{K}$.  $K(X,X)$, $K(X,X_*)$, $K(X_*,X)$, and $K(X_*,X_*)$ are the matrices of covariances between pairs of training and prediction points, the elements of which are calculated using the covariance function $\mathbf{k}(\vec{x},\vec{x} ')$. The conditional probability distribution of $\mathbf{f}_*$ given $\mathbf{f}$ is itself Gaussian (see e.g.~\citep{Rasmussen}):
\beq
\label{eq:conditionalprob}
\begin{aligned}
p(\mathbf{f}_*|\mathbf{f}) = &\mathcal{N}\bigg(K(X_*,X)K(X,X)^{-1}\mathbf{f},\\
& K(X_*,X_*) - K(X_*,X)K(X,X)^{-1}K(X,X_*)\bigg).
\end{aligned}
\eeq
With covariances and a training set, $\mathbf{f}$, one can calculate the conditional mean and covariance of $\mathbf{f}_*$.  In \GPR{}, the entries of the covariance matrices are computed using a covariance function, or kernel, which is specified by the user and depends on the application.  The kernels are symmetric and hence only depend on the distance between points $r=|x-x'|$. Here we primarily consider the squared-exponential kernel, but we discuss the choice of covariance function in \S\ref{sec:discussion}. 
A squared-exponential covariance constrains the process to be infinitely mean-square differentiable\footnote{See \S 4.1.1 in \citealt{Rasmussen} for the definition of mean-square differentiability.} and takes the form of a Gaussian:
 \beq
 \mathbf{k}_{SE}(\vec{x}_1,\vec{x}_2) = \mathbf{k}_{SE}(r=|\vec{x}_1-\vec{x}_2|) = \sigma^2\exp\left(-\frac{1}{2} r^2/l^2\right).\\
 \eeq
where $\sigma$ and $l$ are {\it hyperparameters} of the process and parameterize the signal variance and length scale, respectively. The hyperparameters of this kernel can be fixed a priori, but are typically chosen to maximize the {\it hyperlikelihood}, the likelihood of the training data under the \GP{} prior (Equation \ref{eqn:GPPrior}). Here we instead maximize the {\it hyperposterior} which incorporates prior distributions on the hyperparameters:
\beq{}
\mathrm{hyperposterior} \propto \mathrm{hyperlikelihood}\times \mathrm{hyperprior}.
\eeq{} 
The priors on the hyperparameters, or {\it hyperpriors}, are discussed further in \S\ref{sec:results}. 

In addition to selection of hyperparameters, one must choose small values called ``nuggets"\citep{ANDRIANAKIS20124215} to add to the diagonal of the training points covariance matrix $K(X,X)$ to ensure numerical stability when computing the conditional covariance.  Computation of the conditional covariance can result in a non-positive-semi-definite matrix due to precision errors when the prior covariance matrix has large values off-diagonal. These nuggets are described further in \S\ref{sec:results}.

\subsection{Implementation of \GPR-based models}
\label{subsec:implement}
We now return our focus to the training set projection coefficients $c_i(\params_j)$ at points $\params_j$ in the input parameter space.  To reconstruct waveforms, P14 interpolates the training projection coefficients over the parameter space using a tensor spline interpolation.  We instead use \GPR{} in order to account for uncertainties in the interpolation. Rather than directly interpolate $c_i$, we interpolate {\it regularized} coefficients $\tilde{c}_i$.  The training values $c_i(\{\params_j\}_{j=1}^{n_{\rm train}})$ are regularized by first subtracting a linear fit over the parameter space, then normalizing the residual variance to 1 and removing the mean:
\beq{}
\begin{aligned}
\Delta{c}_i(\{\params_j\})\equiv c_i(\{\params_j\}) - {\rm linfit}\left(c_i(\{\params_j\})\right)\\
\tilde{c}_i(\{\params_j\})\equiv \frac{\Delta{c}_i(\{\params_j\}) - {\rm mean}\left(\Delta{c}_i(\{\params_j\})\right)}{{\rm std}\left(\Delta{c}_i(\{\params_j\})\right)}.
\end{aligned}
\eeq{}
We then assume that each $\tilde{c}_i$ is a Gaussian process, and use $\tilde{c}_i(\{\params_j\})$ as a training set to regress $\tilde{c}_i(\{\params\})$.  Thus for any point in parameter space $\params$ we can predict $\tilde{c}_i(\params)$ and marginal uncertainties $\delta \tilde{c}_i(\params)$ which can then be transformed back to amplitude/phase by applying the reverse of the transformations used to generate $\tilde{c}_i(\{\params_j\})$.  Assuming $c_i$ is uncorrelated with $c_j$ unless $i=j$, the mean amplitude $A(F_k,\params)$ in the $k$-th frequency bin and covariance matrix $\Sigma_{kl}(\params)$ are given by\footnote{The correlations between $c_i$'s are outside the scope of this paper, and will be considered in future work.}:
\beq
\begin{aligned}
A(F_k,\params) = \sum_i V^A_{ki} c^A_{i,{\rm GP}}(\params), \\
\Sigma^A_{kl}(\params) = \sum_i V^{A}_{ki} \delta c^A_{i,{\rm GP}}(\params) V^{A\top}_{il}. \\
\end{aligned}
\eeq    
The phase means and covariances, $\Phi(F_k,\params)$ and $\Sigma^\Phi_{kl}(\params)$, can be reconstructed similarly with the corresponding phase projection coefficients.  

To assess the accuracy of our \GPR{} model, we compare reconstructed frequency waveform means from \GPR{} $h_{\rm \GPR}(F_k,\params)$ with the waveforms from IMRPhenomD $h(F_k,\params)$ using the {\it mismatch} function.  The mismatch between two frequency domain waveforms $h_1$ and $h_2$ is defined as:
\beq
\begin{aligned}
&{\rm mismatch}(h_1,h_2) = \\
&1 - \frac{4}{||h_{1}|| ||h_{2}||} \\
& \times \Re\left(\int_{f_{\rm min}}^{f_{\rm max}}\frac{h_{1}(f) h_2^*(f)}{S_f} {\rm d}f\right),
\end{aligned}
\eeq  
where 
\beq
||h|| = 4\Re\left(\int_{f_{\rm min}}^{f_{\rm max}}\frac{h(f) h(f)^*}{S_f} {\rm d}f\right).
\eeq
$f_{\rm min}$ and $f_{\rm max}$ are the minimum and maximum frequencies at which the two waveforms are compared, respectively, and $S_f$ is the noise power spectral density of a \GW{} detector. In this paper, we use the aLIGO O1 noise curve {\tt 2015-10-01\char`_H1\char`_O1\char`_Sensitivity\char`_strain\char`_asd.txt}\footnote{This noise curve can be found at https://dcc.ligo.org/LIGO-G1501223/public.}.

\section{Results}
\label{sec:results}
We now implement \GPR{}-based models for three sets of parameter spaces using the {\tt GaussianProcessRegressor} module from {\tt scikit-learn}\citep{scikit-learn}:
\begin{enumerate}
\item constant zero spin\\mass ratio $q \in [1, 6]$
\item equal-and-aligned spin $\chi_1=\chi_2 \in [-1,1]$ \\ constant $q=1$
\item equal-and-aligned spin $\chi_1=\chi_2 \in [-0.5,0.5]$ \\ $q \in [1,3]$
\end{enumerate} 
All parameter spaces use waveforms with a constant chirp mass $M_c = 20M_\odot$, distance $D=1$ Mpc, inclination angle $i=0$, and starting frequency of 20 Hz.

 \begin{figure}[ht!]
	\centering
	\includegraphics[scale=0.55]{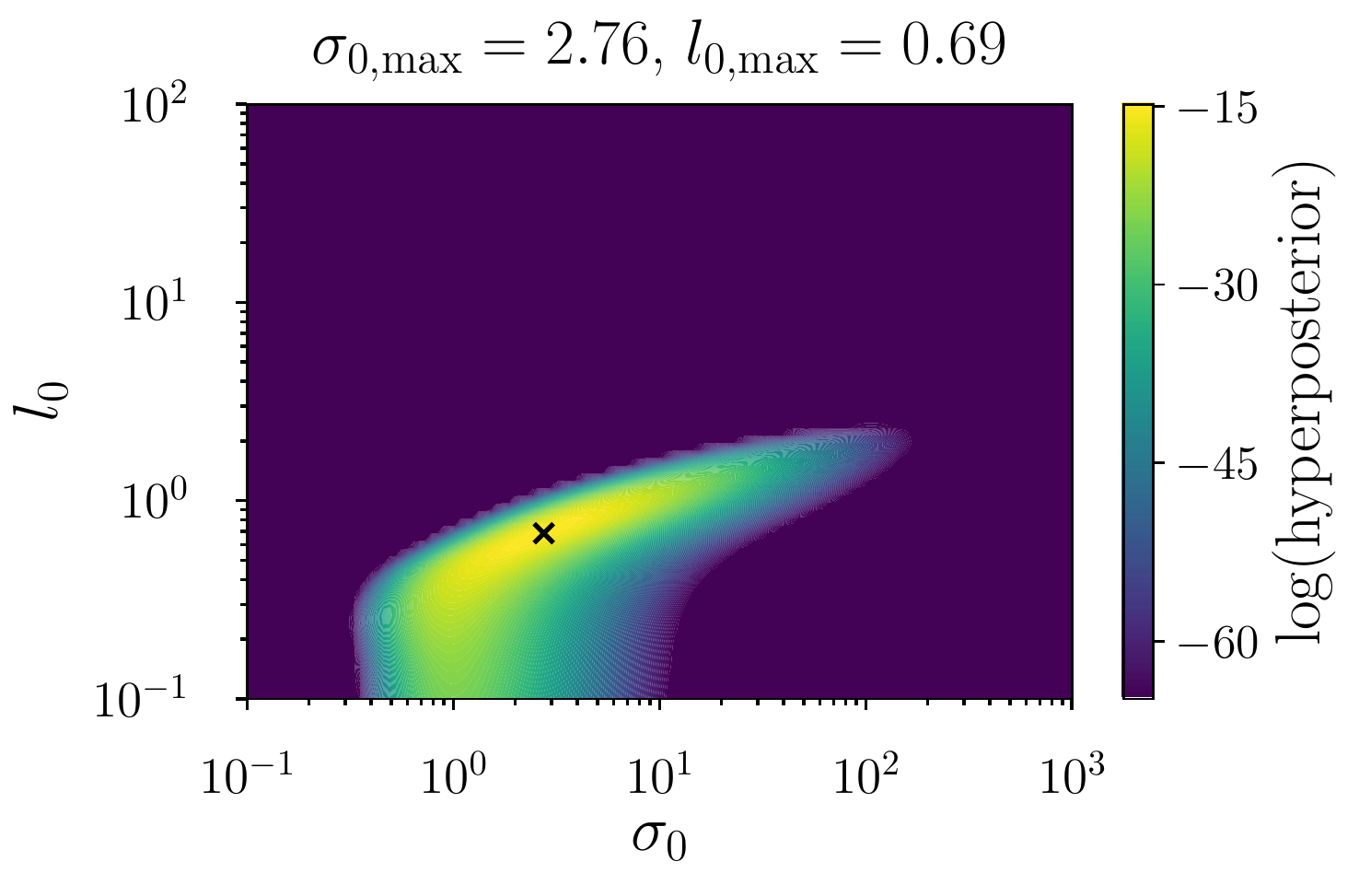}
	\caption{Hyperposterior for the first regularized amplitude coefficient $\tilde{c}_0^A$ as a function of kernel covariance scale $\sigma_0$ and length scale $l_0$.  The hyperparameter values with largest hyperposterior, $\sigma_{0,\rm{max}}$ and $l_{0,\rm{max}}$, are marked with an X on the plot and printed at the top.}
	\label{fig:amp_likelihood}
\end{figure}

\subsection{1-d GPR in Mass Ratio}
\label{sec:GPRq}

Beginning with Parameter Space 1, we generate $n_{\rm train}=15$ equally spaced IMRPhenomD waveforms from $q=1$ to $q=6$ as a proxy for an \NR{} training set and compute their amplitude and phase projection coefficients as described in \S\ref{subsec:WFgen}.  The $i$-th coefficient is then de-trended by removing a linear fit of the 20 training coefficient values $c_i(\{\params_j\})$.  The de-trended $c_i(\{\params_j\})$ are then normalized by their standard deviation and their mean is removed.  We refer to the de-trended, de-meaned, normalized coefficients as $\tilde{c}_i$ and treat each one as a \GP:
\beq
\tilde{c}_i \sim \mathcal{GP}(\vec{0}, \mathbf{k}_i(q,q')).
\eeq
The $\tilde{c}_i$ can be trivially transformed back to $c_i$ by reapplying the mean, standard deviation, and linear fit.  We take the covariance function for the $i$-th coefficient, $\mathbf{k}_i(q,q')$, to be a squared exponential with hyperparameters $\sigma_i$ and $l_i$.  




To ensure numerical stability in the \GPR{} conditional covariance matrix calculation, we add a nugget to the input covariances for each training value.  We assume a constant relative error on the training waveform amplitude of $10^{-4}$ at each frequency and transform these errors to errors in the amplitude coefficients, which are used as the kernel nugget.  For the training phases, the error at each frequency is kept below the nominal LIGO phase measurement uncertainty of $\sim 0.1$ radians by assuming a constant error of 0.01 radians at each phase value on the sparse frequency grid. These errors are projected to the coefficient errors analogously to the amplitude error case.  The nugget levels here are chosen for numerical stability and adequate accuracy, but also roughly correspond to the resolution errors found in \NR{} simulation studies \citep{Lovelace}.  In principle, the \NR{} errors as a function of source parameters could be incorporated into the nugget values to fully account for the resolutions of different simulations.  \\ 
 \begin{figure}
	\centering
	\includegraphics[scale=0.34]{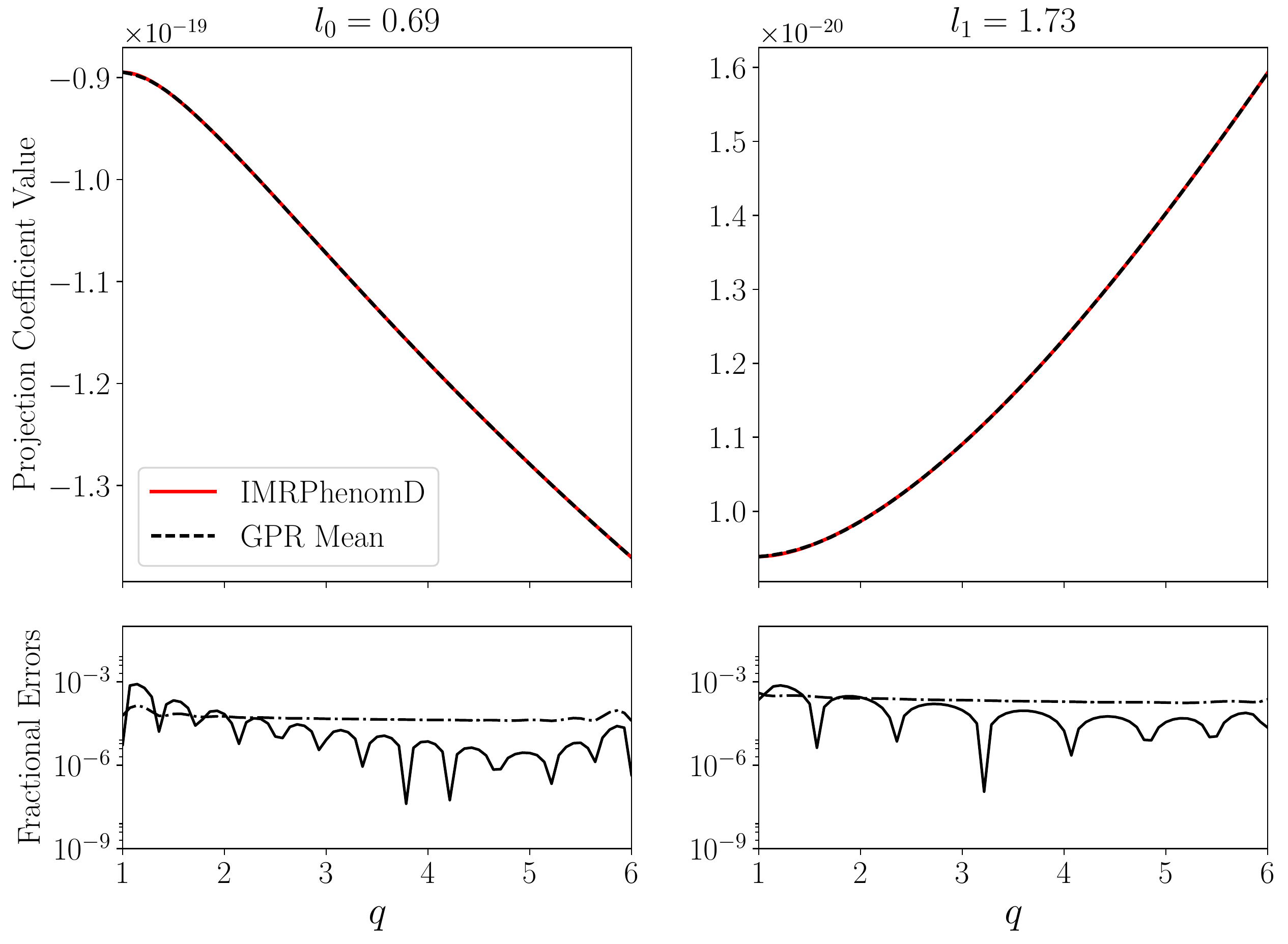}
	\caption{First two amplitude coefficients for the IMRPhenomD waveforms and \GPR-based waveforms with their residuals. {\it Top panels}: IMRPhenomD coefficient values (red) and \GPR{} mean coefficient values (black, dashed) for the first two amplitude coefficients. The optimized length scales are shown above the top panels. Although the remaining coefficients are not shown here, they have similar morphologies. {\it Bottom panels}: the fractional residuals $(|c^A_i - c^A_{i,{\rm GP}}|)/{c^A_i}$ (solid) and the \GPR{} fractional 1$\sigma$ uncertainties $\delta c^A_{i,{\rm GP}}/{c^A_i}$ (dashed-dotted).}
	\label{fig:amp_coeffs}
\end{figure}
 \begin{figure}
	\centering
	\includegraphics[scale=0.34]{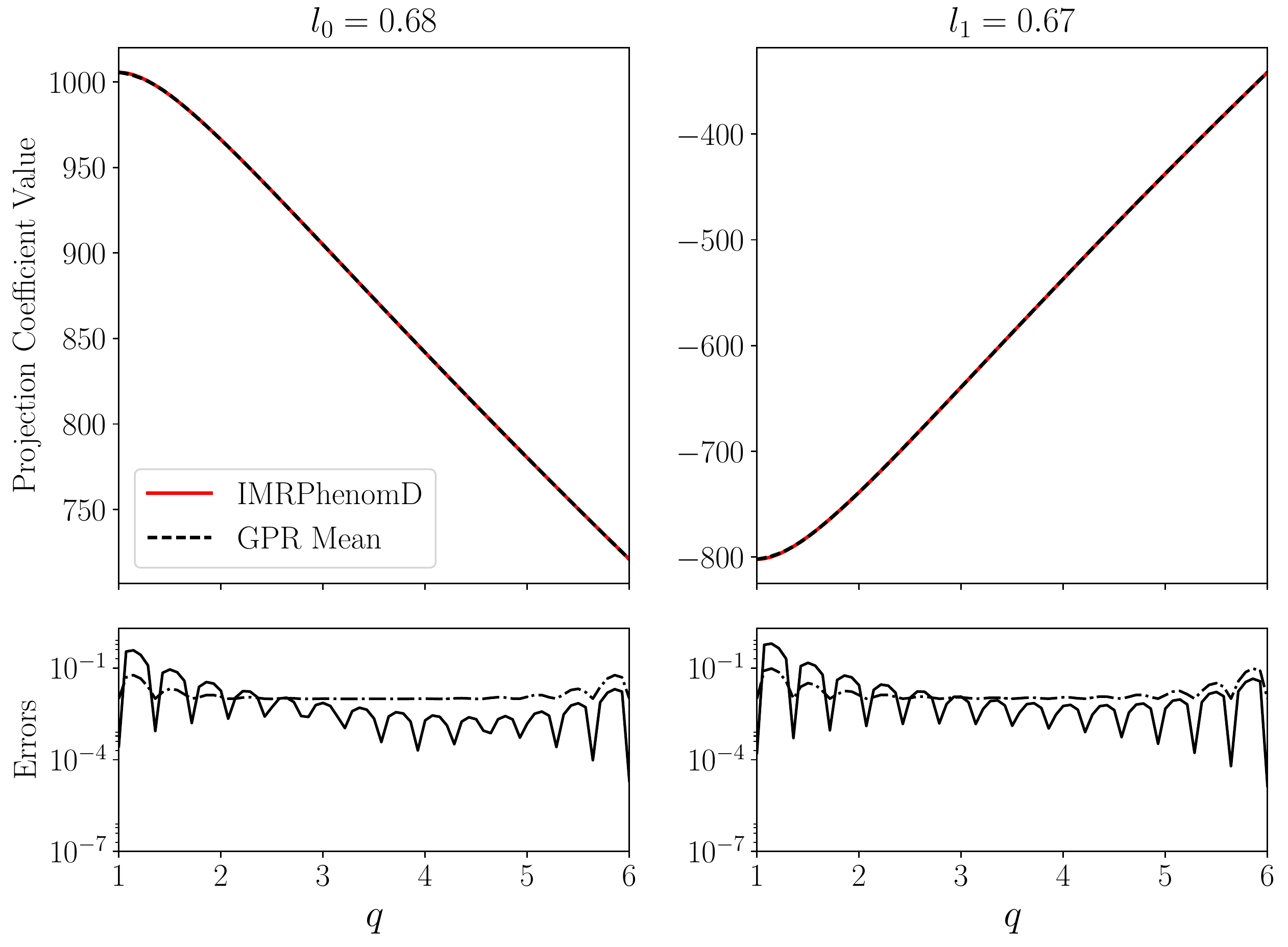}
	\caption{First two phase coefficients for the IMRPhenomD waveforms and \GPR-based waveforms with their residuals. {\it Top panels}: IMRPhenomD coefficient values (red) and \GPR{} mean coefficient values (black, dashed) for the first two phase coefficients. The optimized length scales are shown above the top panels. Although the remaining coefficients are not shown here, they have similar morphologies. {\it Bottom panels}: the residuals $|c^\Phi_i - c^\Phi_{i,{\rm GP}}|$ (solid) and the \GPR{} 1$\sigma$ uncertainties $\delta c^\Phi_{i,{\rm GP}}$ (dashed-dotted).}
	\label{fig:phase_coeffs}
\end{figure}
To optimize the kernel hyperparameters for each coefficient, we use the {\tt scipy.optimize} implementation \citep{scipy} of the Broyden-Fletcher-Goldfarb-Shanno algorithm {\tt fmin\char`_l\char`_bfgs\char`_b} \citep{Zhu} to maximize the hyperposterior. We apply log-normal hyperpriors on $\sigma_i$ and $l_i$:
\beq{}
\begin{aligned}
&\log_{10}(\sigma_i)\sim \mathcal{N}(0, 0.5)\\
&\log_{10}(l_i)\sim \mathcal{N}\bigg(\log_{10}\left(\frac{1}{2}{\rm width}(\{q_j\}_1^{n_{\rm train}})\right),1\bigg).
\end{aligned}
\eeq{}
Since the regularized coefficient functions being interpolated have been normalized by their standard deviation, we expect that $\sigma_i\sim 1$, motivating the hyperprior on $\sigma_i$ above.  We have less information about the length scales a priori, but we know that they should not be much shorter than the distance between the closest training points, nor should they be much larger than the width of the parameter space. As such, the normal distribution on $\log_{10}(l_i)$ is chosen to have a standard deviation of 1 (i.e. 1 order of magnitude) and peak at half the width of the parameter space spanned by the training set. Figure \ref{fig:amp_likelihood} shows the hyperposterior surface for the first amplitude coefficient as a function of the hyperparameters using the squared-exponential kernel. The hyperposteriors for other coefficients are similar in morphology to those shown here. 

 \begin{figure}[tbp!]
	\centering
	\includegraphics[scale=0.53]{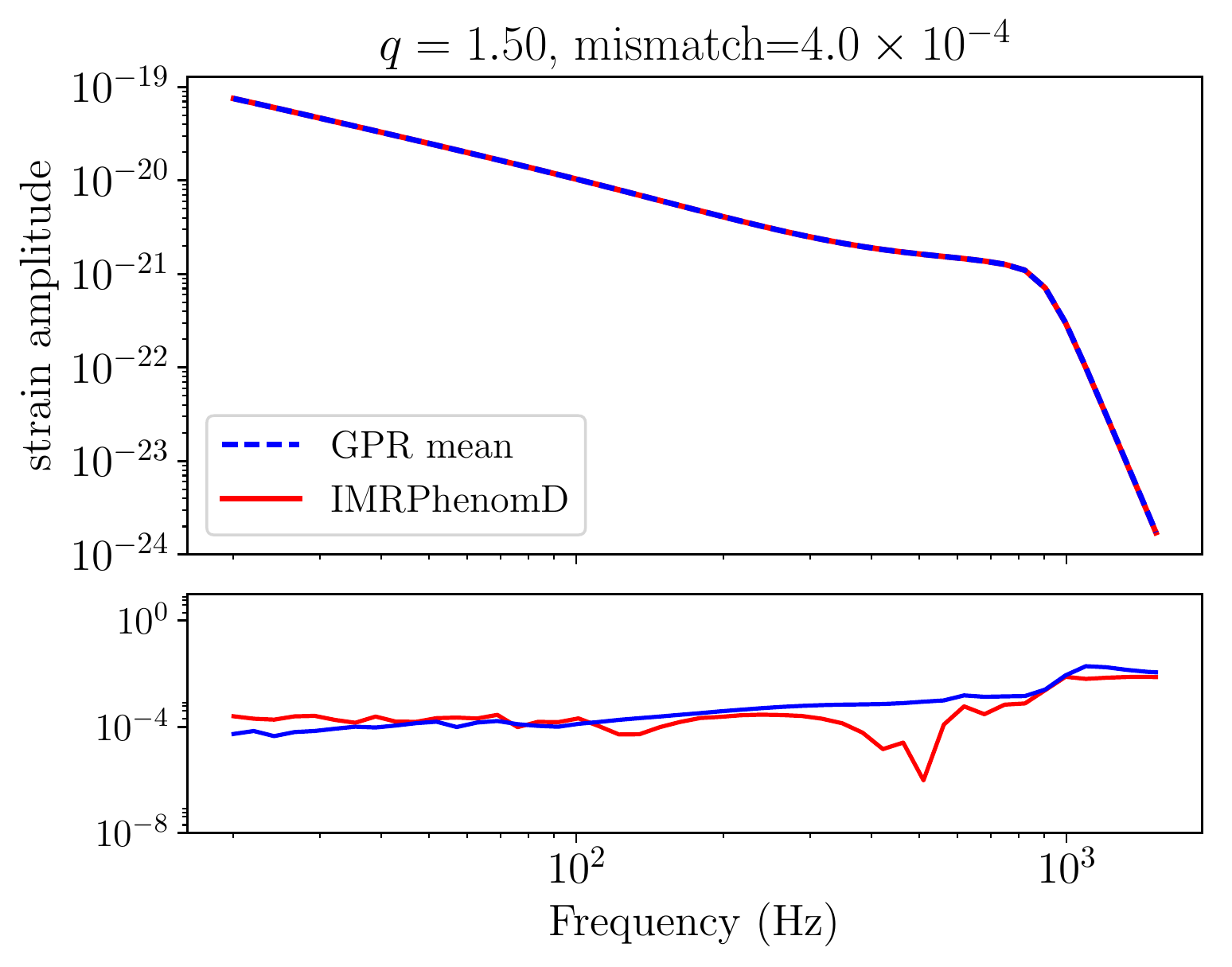}
	\caption{Example reconstructed \GPR{} amplitude function.  {\it Top:} The interpolated \GPR{} mean amplitude vs. frequency is shown in blue and the IMRPhenomD amplitude is overlaid in red. {\it Bottom:} The IMR-GPR-mean residual amplitude is shown in red, and the \GPR{} 1$\sigma$ uncertainty is shown in blue as a function of frequency. Both errors are normalized to the IMRPhenomD amplitude.}
	\label{fig:amp_sample}
\end{figure}

 \begin{figure}[tbp!]
	\centering
	\includegraphics[scale=0.53]{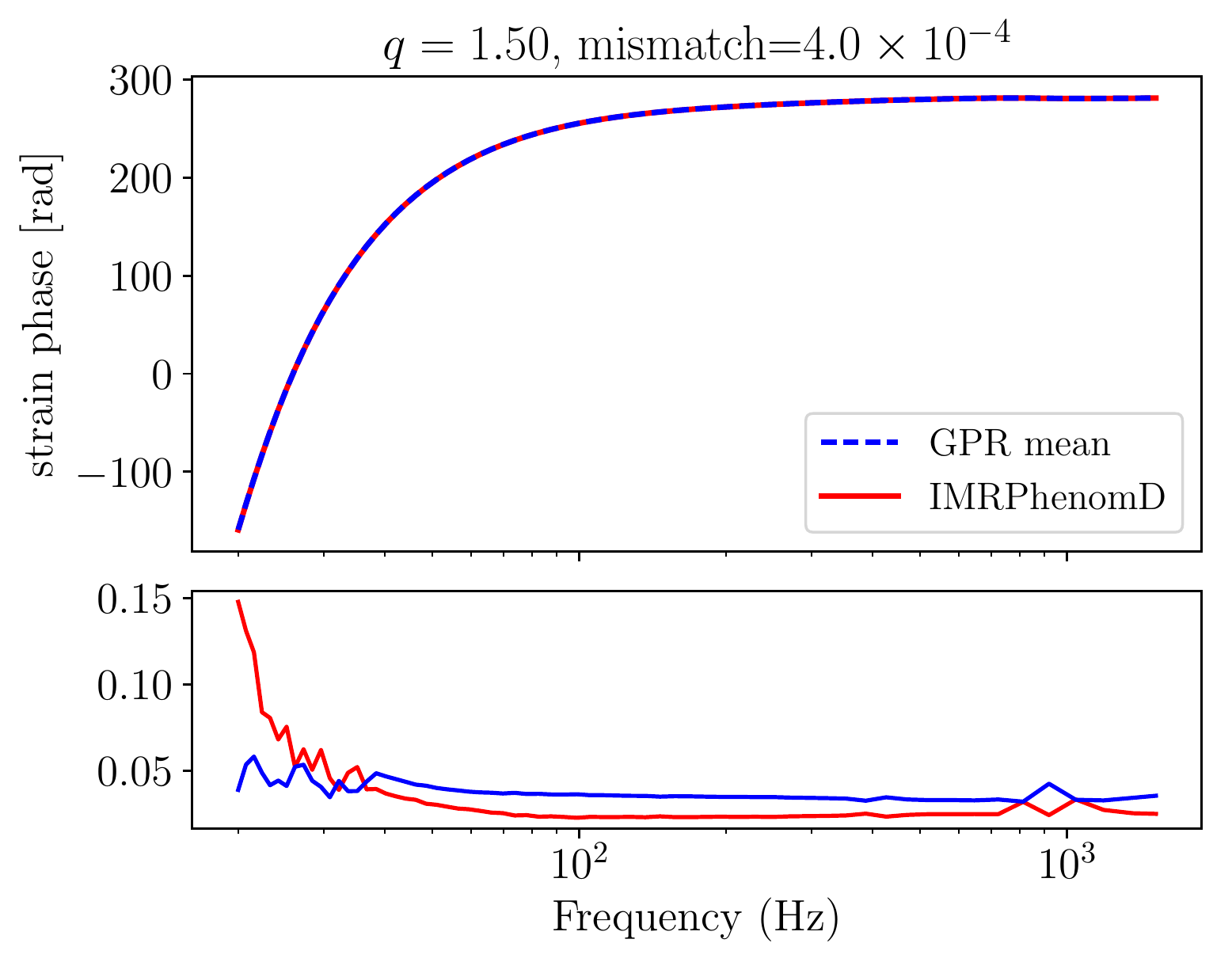}
	\caption{Example reconstructed \GPR{} phase function.  {\it Top:} The interpolated \GPR{} mean phase vs. frequency is shown in blue and the IMRPhenomD phase is overlaid in red. {\it Bottom:} The absolute value of the IMR-GPR residual phase is shown in red, and the \GPR{} 1$\sigma$ uncertainty is shown in blue as a function of frequency.}
	\label{fig:phase_sample}
\end{figure}

\begin{figure}
	\centering
	\includegraphics[scale=0.55]{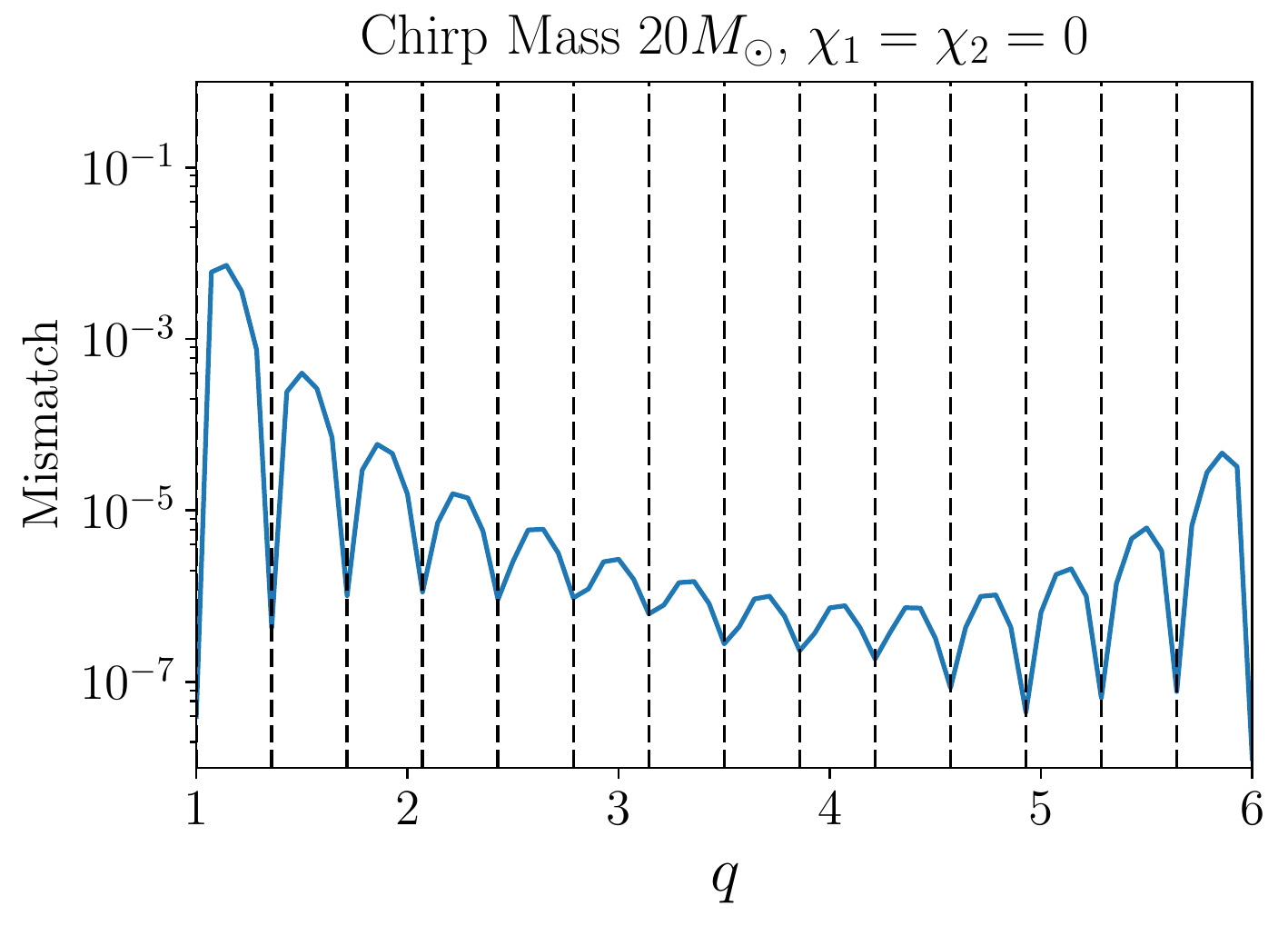}
	\caption{Mismatch between the IMRPhenomD waveform and the \GPR{} mean waveform for different mass ratios assuming a constant chirp mass and zero spin. The dashed, vertical lines show the IMRPhenomD training waveform locations used to build the \GPR{} model.}
	\label{fig:mismatch_q}
\end{figure}

With optimized hyperparameters, the \GP{} is used to interpolate each regularized coefficient on a grid five times finer than the training grid.  With the \GPR{} model of the $\tilde{c}_i$'s, we can calculate each $c_i$ and transform them back to amplitudes and phases on the sparse frequency grid.  Figures~\ref{fig:amp_coeffs} and~\ref{fig:phase_coeffs} show the values of the first two amplitude and phase $c_i$'s, respectively, as a function of $q$ from IMRPhenomD and from the \GPR-based model. Also shown are the fit residuals, the \GPR{} $1\sigma$ uncertainties, and the optimized length scale hyperparameter for each coefficient function's \GP{}.  Although the \GPR{} errors do not perfectly match the residuals across the parameter space, they are indicative of the maximum error level and of the fact that the errors are largest on the edges of the space.  
The interpolated coefficients and their uncertainties are then propagated back to amplitudes and phases on the sparse frequency grid.  One such example of a \GPR-interpolated waveform is shown in Figures \ref{fig:amp_sample} and \ref{fig:phase_sample}, which show the amplitude and phase functions, respectively. Notably, the fractional amplitude error between the \GPR{} model and IMRPhenomD waveform is largest at high frequencies, because small errors in the amplitude coefficients combine when projected back to the sparse frequency grid domain.  Nonetheless, the \GPR{} mean agrees well with the IMRPhenomD model, and the \GPR{} uncertainties give a reasonable indication of the true error levels at different frequency bins.  Since \PE{} is done in the frequency domain, these errors do not need to be propagated to the time domain, although frequency-domain waveforms can be sampled to create a distribution of time-domain waveforms if required.
Figure \ref{fig:mismatch_q} shows the mismatch between the \GPR{} mean waveform and the IMR waveform at various values of $q$.  
The mismatch increases near the boundaries of the space where there are fewer nearby training points (training points shown in dashed black vertical lines).  Indeed, the mismatches and residuals suggest that more training points (i.e.~simulations) are needed towards the edge of the space of interest.  Increasing the number of training points typically lowers the mismatches, although it depends on the nugget and hyperparameters used in the \GP. The relative placement of the simulations is also of interest, which we discuss in \S\ref{sec:optimize}.  


\subsection{1-d GPR in equal-and-aligned spin}
\label{sec:GPRspin}
\begin{figure}
	\centering
	\includegraphics[scale=0.55]{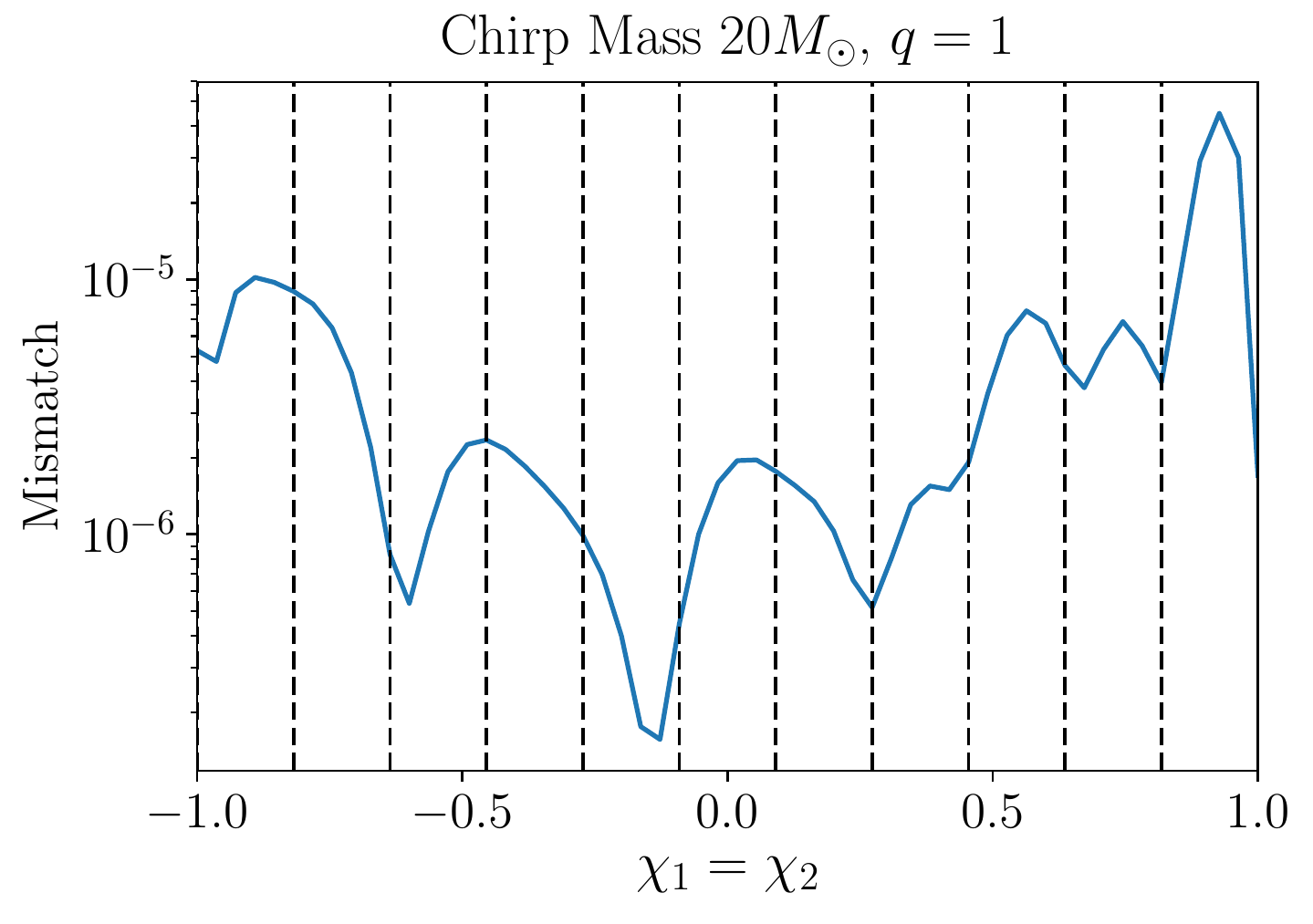}
	\caption{Mismatch between the IMRPhenomD waveform and the \GPR{} mean waveform for different equal-and-aligned spin values assuming a constant chirp mass and equal masses ($q=1$). The dashed, vertical lines show the IMRPhenomD training waveform locations used to build the \GPR{} model.}
	\label{fig:mismatch_chi}
\end{figure}

Following an analogous method to that presented in \S\ref{sec:GPRq}, we generate 12 IMRPhenomD waveforms spanning equally-spaced equal-and-aligned spin\footnote{The spins of the inspiraling objects are assumed to be equal and aligned with the orbital angular momentum.} values from $\chi=-1$ to $\chi=1$. The chirp mass of $M_c=20M_\odot$ and mass ratio $q=1$ are held fixed. A \GPR-based waveform model is built using these waveforms as a training set, again using a squared-exponential covariance function to model each $\tilde{c}_i$.  The hyperpriors on the length scale and covariance scale are chosen in the same way as in \S\ref{sec:GPRq}. 
The \GPR{} model is evaluated on a grid five times finer than the training grid and is then compared to the waveforms predicted in IMRPhenomD via the mismatch function.  Figure \ref{fig:mismatch_chi} shows the mismatch between the \GPR{} mean waveform and IMRPhenomD for the 1-d space of equal-and-aligned spin $\chi$.  
Similarly to the case of mass-ratio space, the mismatch between the \GPR{} model and IMRPhenomD is largest at the boundaries of the training set.  Nevertheless, we are able to achieve low mismatches with just a few training points in this case.

\subsection{2-d GPR in equal aligned spin and mass ratio}
\label{sec:2DGPR} 

\begin{figure*}[ht!]
	\centering
	\includegraphics[scale=0.6]{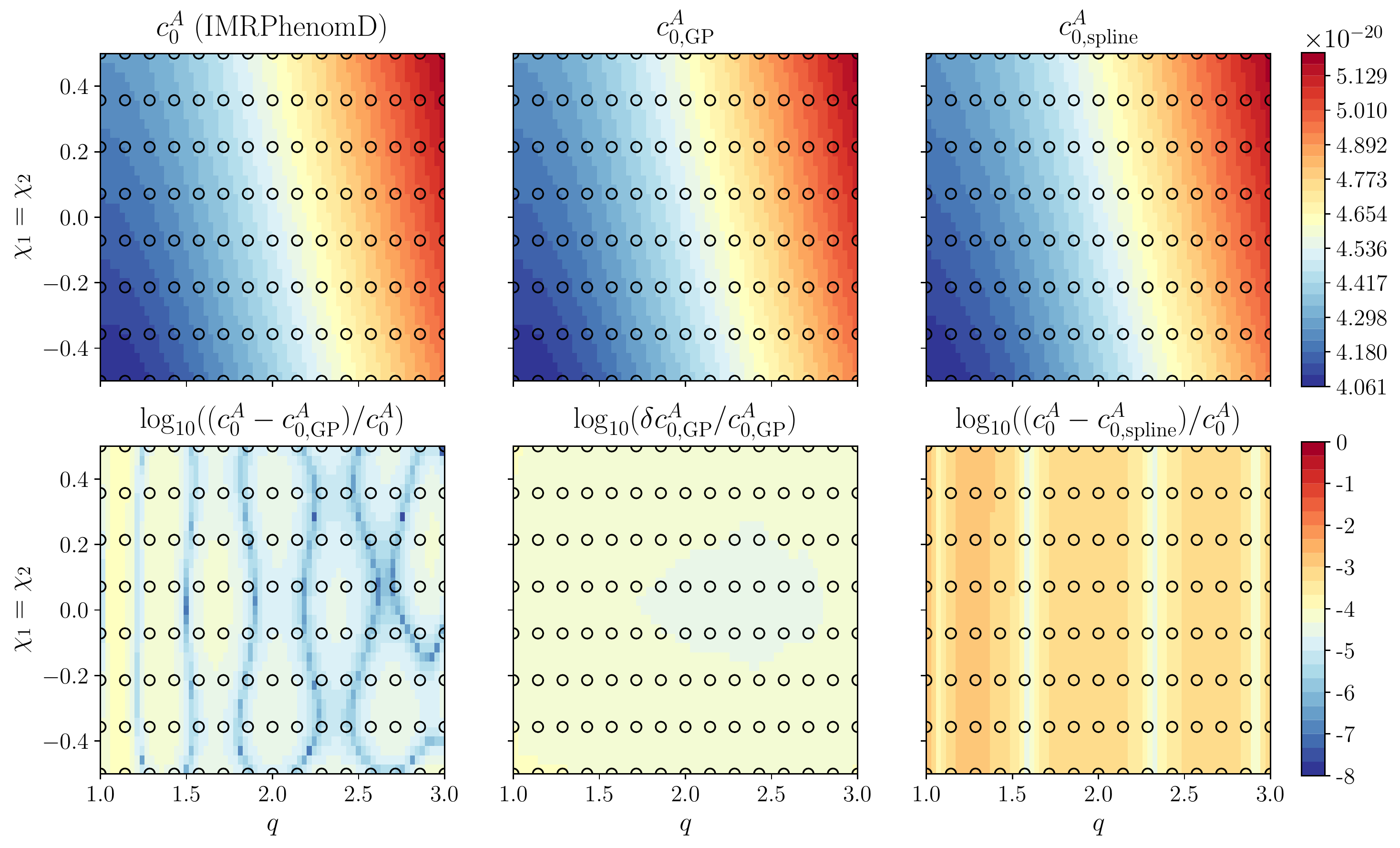}
	\caption{First amplitude coefficient $c_0^A$ as a function of $q$ and equal-and-aligned spin $\chi_1=\chi_2$. Black circles show the training point locations.  {\it Top left}: $c_0^A$ from the accurate model IMRPhenomD from which the training points are generated.  {\it Top middle}: The \GPR{} mean interpolation of $c_0^A$. {\it Top right}: The B-spline interpolation of $c_0^A$. {\it Bottom left}: The log of the fractional residual of $c_0^A$ between IMRPhenomD and the \GPR{} mean.  {\it Bottom middle}: the log of the fractional $1\sigma$ uncertainty on $c_0^A$ from the \GPR{}.  {\it Bottom right}: The log of the fractional residual of $c_0^A$ between IMRPhenomD and the B-spline.}
	\label{fig:coef_2D}
\end{figure*}

We again build a training set using IMRPhenomD waveforms as a proxy for \NR{} simulations in order to train a \GPR-based model, except here we vary two source parameters: The mass ratio and the value of the equal-and-aligned spins.  The training waveforms are generated on a regular grid with 15 points in $q$ and 8 points in $\chi$ from $q\in [1,3]$ and $\chi \in [-0.5,0.5]$.  The kernel used here is simply an overall covariance $\sigma$ times the product of two squared-exponential kernels, one for the $q$ dimension and one for the $\chi$ dimension.  As such, there are now two length scale parameters, $l_q$ and $l_\chi$, as well as the overall covariance $\sigma$ which need to be fit for each coefficient function.  Hyperpriors for the hyperparameters are chosen as in previous sections, and the hyperparameters are optimized via the hyperposterior. For computational savings, smaller windows in $q$ and $\chi$ are considered here than in the previous sections.  Extending to the full parameter space would simply require more training points\footnote{The computational time to build a \GPR{} given hyperparameters and $N$ training waveforms goes as $\sim O(N^3)$.  Additional cost is incurred by the hyperposterior maximization.}. Another option would be to decompose the extended domain into smaller overlapping patches and build a \GPR{} model for each patch (e.g. Figure 2 of \citep{Purrer2016}).  Although a regular, equal-spacing grid leads to reasonable mismatches in this case, there is no a priori reason to use such a grid, and in practice the existing simulations will have non-regular placements throughout the parameter space. 
Figure \ref{fig:coef_2D} shows the ``true" values, interpolations, and residuals of the first amplitude SVD coefficient $c^A_0$ as a function of $q$ and $\chi$.  The top left panel is a color map of the coefficient values from IMRPhenomD.  The top middle and right panels show $c^A_{0,{\rm GP}}$ and $c^A_{0,{\rm spline}}$, which are the interpolations of $c^A_0$ with \GPR{} and B-splines, respectively. The bottom panels show the fractional residuals of the interpolants and the estimated error from \GPR{}. The residuals between IMRPhenomD and the \GPR{} model (bottom left panel) for this amplitude coefficient are below the $0.1\%$ level and are comparable to the predicted \GPR{} uncertainties, providing evidence for the accuracy and precision of the \GPR{} model.  Comparing the bottom left and bottom right panels of Figure \ref{fig:coef_2D}, the \GPR{} mean is roughly as accurate at the B-spline interpolation for most coefficients, but the spline does not give any information about interpolation errors.  The bottom middle panel shows the \GPR{}-estimated fractional errors which give an estimate of the maximum true \GPR{}-IMR residual.
 \begin{figure}[htbp!]
	\centering
	\includegraphics[scale=.6]{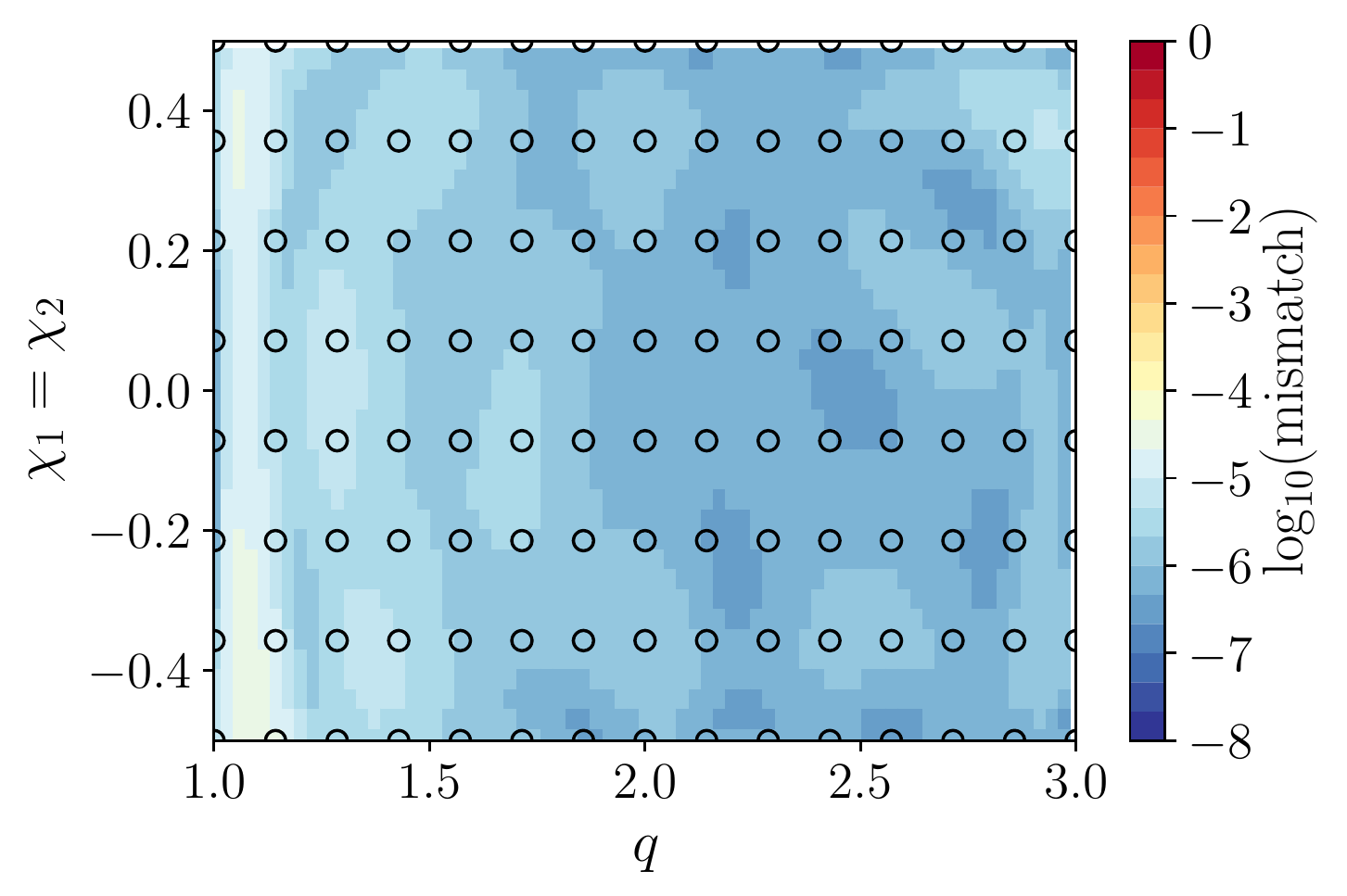}
	\caption{Mismatch between IMRPhenomD waveforms and \GPR{} mean waveforms with a regularly-gridded training set.  The black circles show the locations of training waveforms from IMRPhenomD used to train the \GPR{}.  There are $15\times8=120$ training points on this grid, and the maximum mismatch in the region is $4.3\sn{-3}$.}
	\label{fig:mismatch2D}
\end{figure}

Figure \ref{fig:mismatch2D} is analogous to the mismatch plots shown in \S\ref{sec:GPRq} and \S\ref{sec:GPRspin}, but now in two dimensions.  The black circles denote the IMRPhenomD training points, and the color map shows the mismatch between the IMRPhenomD waveform and the \GPR{} mean waveform at each point. The mismatches calculated here are all at or below $4.3\sn{-5}$.  

\section{Where should we run the next numerical relativity simulation?}
\label{sec:optimize}
Two natural questions arise from considering a \GP{} model: 1) what is the error level of the \GPR{} model for different GW source parameters, and 2) given a pre-existing set of training waveforms, where is the ``optimal" placement of an additional training waveform?  Thus far, we have evaluated the accuracy of our \GPR{} models by computing the mismatch between the \GPR{} model and the IMRPhenomD, but in practice such comparisons will not be possible, since the ``true waveform" (i.e.~\NR{} simulation results) will be unknown everywhere other than at the existing simulation points.  As such, the mismatch between the \GPR{} mean and the true waveform cannot be used to determine the \GPR{} error level nor can it be used as a parameter of the optimization function which selects new \NR{} simulation parameters.  Instead, we propose using the \GPR{} posterior uncertainties to guide the overall error estimation and training set optimization.  
Here we present a simple metric for estimating \GPR{} waveform errors and for choosing where in parameter space to add a new training waveform.  The basic idea is to estimate the mismatch between a \GPR{} and \NR{} waveform with the same parameters based on the spread of \GPR{} samples.  Specifically, we estimate the \GPR{}-IMR mismatch by computing the largest mismatch between $M$ \GPR{} samples and the \GPR{} mean. New accurate waveforms can be added where this estimated mismatch is large, analogously to the greedy algorithm 8.1 in \citep{Rasmussen}.  We summarize this training point placement strategy in Algorithm \ref{alg:iterative}.

\begin{algorithm}[H]
\caption{greedy training point placement}
\label{alg:iterative}
\begin{algorithmic}
\State{$\{\params_j\} \gets n_{\rm train}$ initial parameter values, $j\in[1,\ntrain]$}
\State{$\{\params^*_k\} \gets$ fine interpolation grid, $k\in[1,\ninterp]$}
\Loop
	\State{Calculate regularized coefficients $\tilde{c}_i(\{\params_j\})$}
	\State{$\tilde{c}_i(\params^*_k) \sim \textrm{GPR}(\tilde{c}_i(\{\params_j\}))$}
	\For{$k \in [1,\ninterp]$}
		\State{$m \gets 0$, $O_k \gets 0$}
		\For{$m\in[1,M]$} 
			\State{$O \gets \texttt{mismatch}(h_{\rm \GPR{}}^{\rm mean}(\params^*_k),h_{\rm \GPR{}}^{\rm sample}(\params^*_k)) $} 
			\State{$O_k \gets \max(O \cup O_k)$}	
		\EndFor
	\EndFor
	\State{$\params_{\ntrain+1} \gets \params^*[{\rm argmax}_k(O_k)]$}
	\State{$\{\params_j\} \gets \{\params_j\} \cup \{\params_{\ntrain+1}\}$}
	\State{$\ntrain \gets \ntrain +1$}
\EndLoop
\end{algorithmic}
\end{algorithm}

First, a few training waveforms, preferably on the boundaries of the parameter space $P$, are used to seed a \GPR{} model of the waveforms in $P$.  \GPR{} waveforms are interpolated on a fine grid in $P$, and at each grid point the maximum mismatch between the \GPR{} waveform mean and $M$ \GPR{} waveform samples is recorded (hereafter called $O_k$ for the $k$-th interpolation point).  $O_k$ at each interpolation grid point is used as a proxy for the true mismatch between the \GPR{} mean and \NR{} in order to determine where to generate a new simulation.  By adding a new simulation to the training set at the point with largest $O_k$, the greedy algorithm attempts to minimize error in locations in parameter space with the largest estimated error.  

\begin{figure}
	\centering
	\includegraphics[scale=.6]{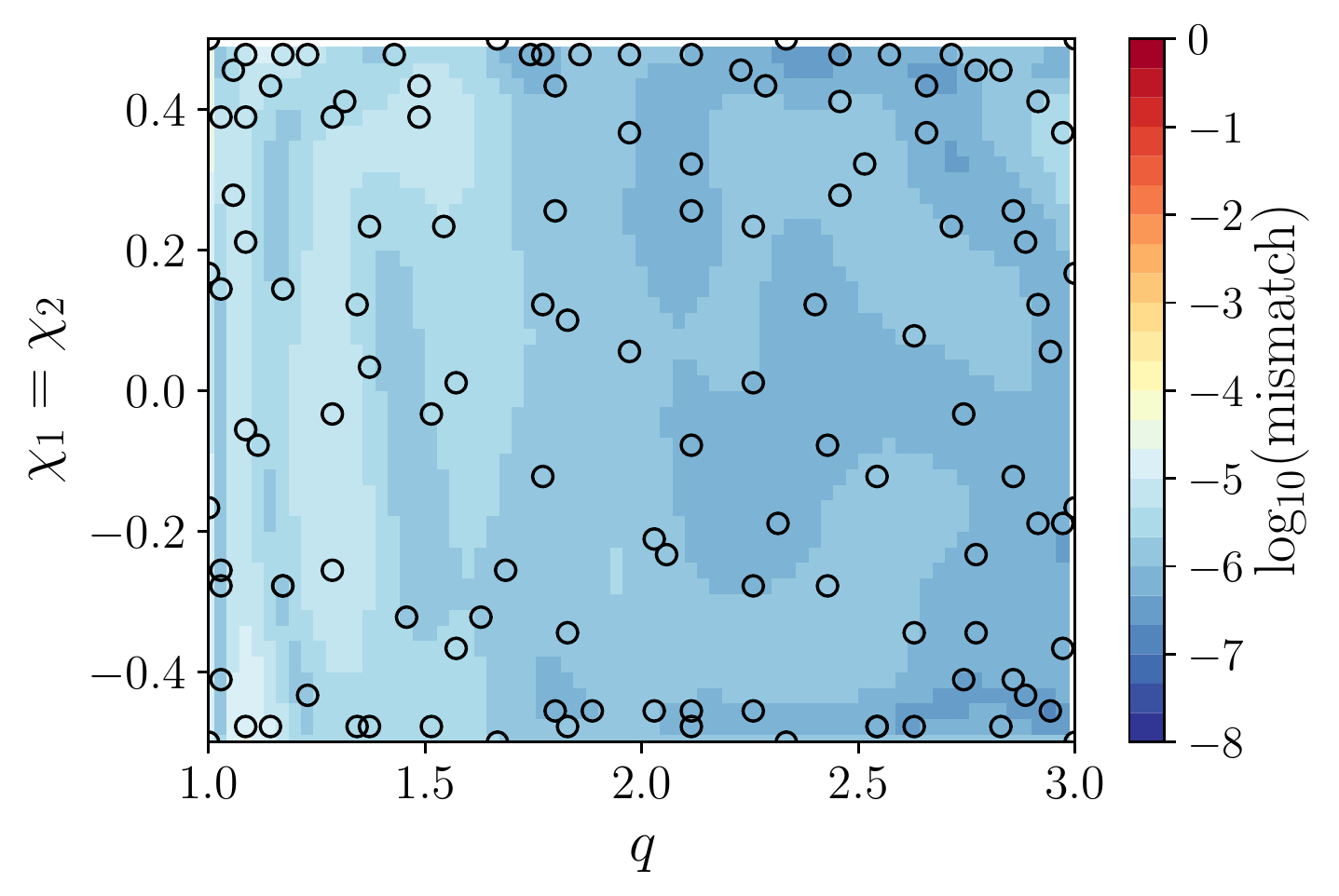}
	\caption{Mismatch between IMRPhenomD waveforms and \GPR{} mean waveforms with an iteratively-built training set.  The training set shown here was seeded with 12 initial IMRPhenomD waveforms and 10 points were added at each iteration based on 100 samples of $O_k$ across the space.  The black circles show the locations of training waveforms from IMRPhenomD used to train the \GPR{}.  In this example, 10 iterations were performed, yielding $10\times10+12=122$ training points and a maximum mismatch in the region of $3.4\sn{-5}$.}
	\label{fig:mismatch2D_iter}
\end{figure}

As a proof of concept, we apply a computationally simplified variant of the greedy algorithm to \GPR{} interpolations in the same $q$--$\chi$ space as in \S\ref{sec:2DGPR}.  Rather than determining $O_k$ at every point on a dense interpolation grid, we instead partition the space into 100 equally-sized, rectangular domains and determine $O_k$ at a random point in each domain.  These 100 $O_k$ values are then used to determine training point placement.  Additionally, at each iteration we add a training waveform at the ten points with highest $O_k$ of the 100 computed, rather than just adding one at a time.  
In the example we show here, we seed the \GPR{} model with 12 initial waveforms: one on each corner of the parameter space and two equally-spaced training waveforms on each edge.  We perform 11 iterations (i.e. 122 total training points) of the greedy algorithm, and compute the mismatch between the IMR model and the \GPR{} mean just as in \S\ref{sec:2DGPR}.  Figure \ref{fig:mismatch2D_iter} shows these mismatch values over the parameter space.  Comparing Figure \ref{fig:mismatch2D_iter} to Figure \ref{fig:mismatch2D}, which have 122 and 120 training points, respectively, we see that the iterative method results in lower mismatches than the regular grid across the parameter space.  Additionally, the maximum mismatch over the parameter space in the iterative case is $9.3\sn{-5}$, which is an order of magnitude lower than the maximum mismatch over the regular grid of $1.4\sn{-3}$. 

 \begin{figure}
	\centering
	\includegraphics[scale=.6]{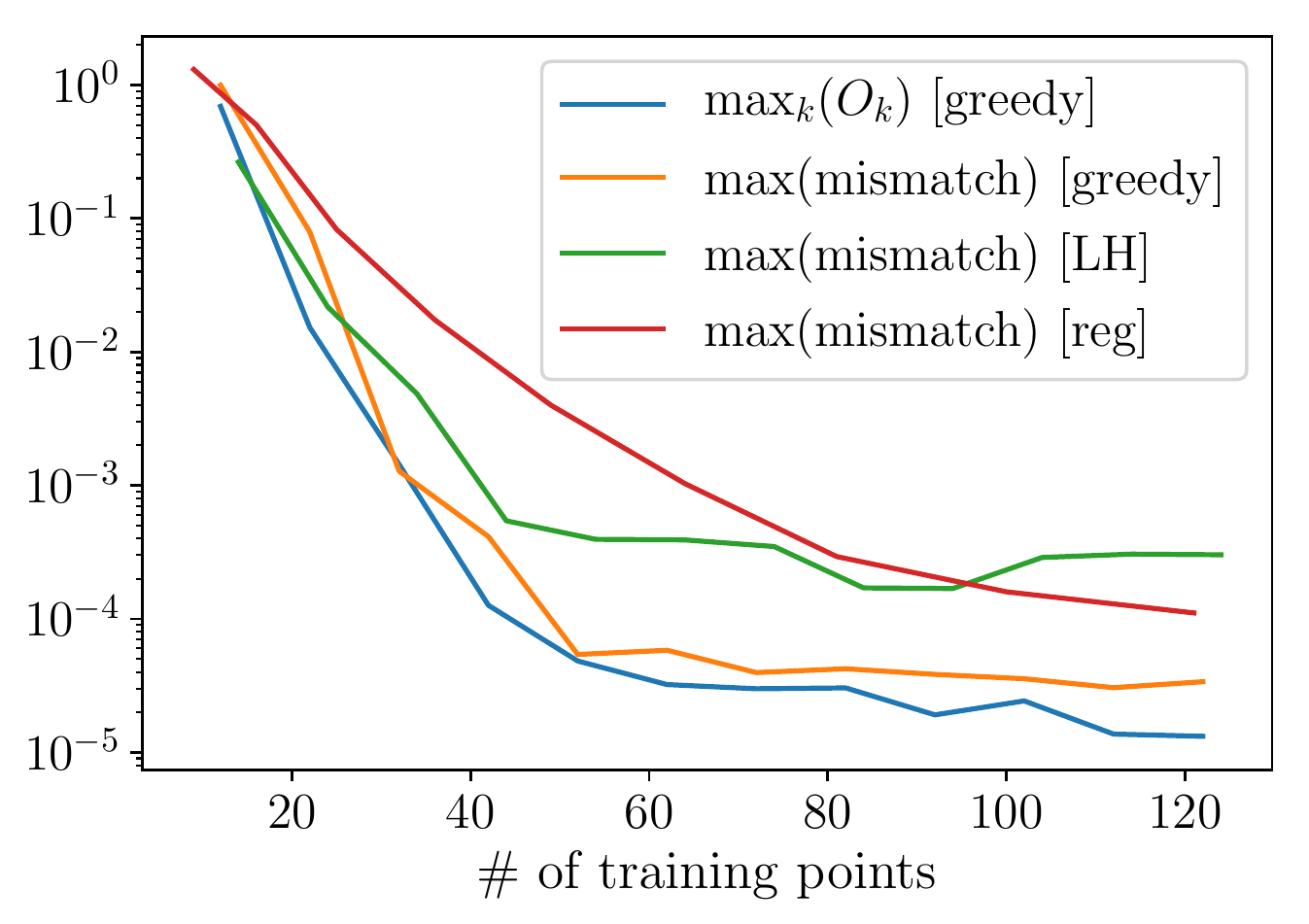}
	\caption{The maximum $O_k$ value using the greedy algorithm and maximum mismatch between the GPR mean and IMRPhenomD for different training point schemes and numbers of training points.  For the greedy training point placement, twelve training waveforms on the boundaries of the space from IMRPhenomD seed the GPR model at the first iteration.  At each subsequent iteration, the ten points with the highest $O_k$ (out of 100 points tested) decide the locations for new training waveforms.  The maximum $O_k$ with greedy placement is shown in blue, and maximum mismatches for greedy, Latin-hypercube, and square grids are shown in orange, green, and red, respectively.}
	\label{fig:error_iter}
\end{figure}

\begin{figure}
	\centering
	\includegraphics[scale=.6]{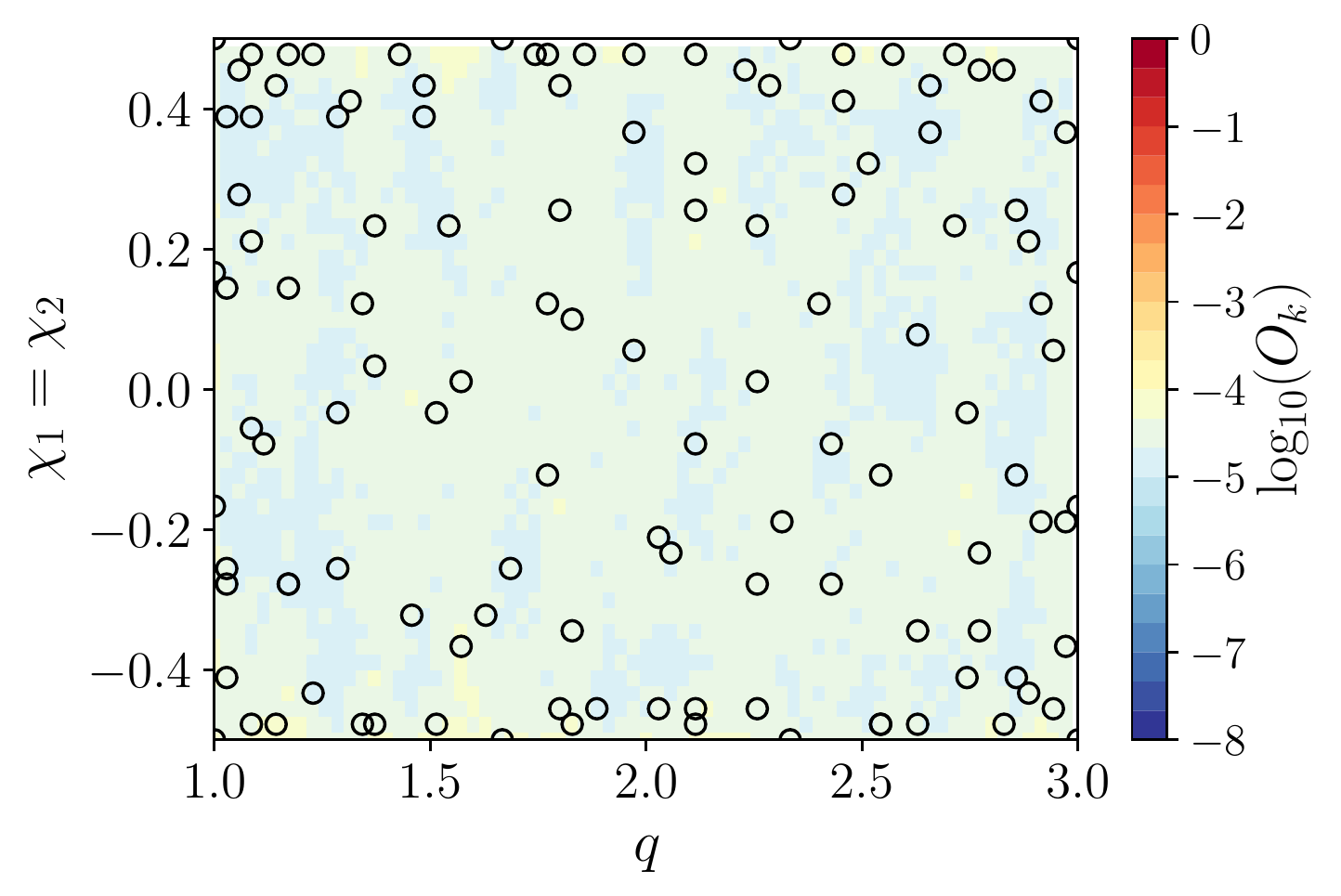}
	\caption{Maximum \GPR{} sample-mean mismatch $O_k$ over 20 samples calculated at each point on the fine interpolation grid based on the same iteratively-built training set shown in Figure \ref{fig:mismatch2D_iter}. The black circles show the locations of training waveforms from IMRPhenomD used to train the \GPR{}.  The maximum $O_k$ in the region is $9.3\sn{-5}$.}
	\label{fig:mismatch2D_est}
\end{figure}

To address the question of whether we can estimate the true mismatches using $O_k$, we compare the maximum of the 100 $O_k$ values and the maximum \GPR{} mean-IMR mismatch over the parameter space at each iteration in Figure \ref{fig:error_iter}.  Using a greedy grid, the maximum $O_k$ value (blue) tracks the maximum \GPR{} mean-IMR mismatch (orange) to within an order of magnitude, though the maximum $O_k$ would likely be higher if more than 100 $O_k$ samples were taken. On the last iteration, we calculate $O_k$ on the fine interpolation grid used in~\S\ref{sec:2DGPR} rather than just sampling 100 $O_k$ values.  This yields the $O_k$ map shown in Figure \ref{fig:mismatch2D_est}.  By construction, $O_k$ is relatively constant over the parameter space.  Additionally, the maximum $O_k$ on this finer grid is $9.3\sn {-5}$, which bounds the maximum \GPR-IMR mismatch of $3.4 \sn {-5}$, further suggesting that the maximum $O_k$ value can be used to estimate the maximum true error level of the \GPR{} mean. As such, $O_k$ can indicate when a sufficient number of training waveforms have been used.

It is worth noting that the maximum mismatch of $4.3 \sn {-5}$ on the 120-point regular grid from \S\ref{sec:2DGPR} is comparable to the maximum mismatch of $3.4 \sn {-5}$ using the 122-point greedy grid.  However, this fact does not indicate that regular training grids are as effective as greedy grids: On the regular grid, the number of training points in the $q$-direction (15) and in the $\chi$-direction (8) were tuned to achieve low mismatches.  In practice though, such tuning will not be possible since (a) the true \GPR-mean error will not be known at points without simulations, and (b) building different regular grids for tuning would use significant simulation resources.

To compare the greedy algorithm to other training point placement schemes, Figure \ref{fig:error_iter} also shows the maximum mismatch over the $q$-$\chi$ space between \GPR{} and IMR as a function of the number of training points for a Latin hypercube (LH) grid (green) and a regular, square grid (red).  For the LH case, a training point is placed at each corner of the space, and then the space is LH sampled with multiples of 10 additional training points.  In other words, for each trial, training points are put on the corners of the space, and a new hypercube with $10\times n$ partitions per axis ($n\in [1,12]$) is constructed and randomly populated with training points under the constraint that there is exactly one training point in each row and column of the hypercube.  For the square grid case, an equally-spaced $n\times n$ training grid spanning the space of interest is created for $n\in[3,11]$. Examining Figure \ref{fig:error_iter}, the greedy algorithm is able to achieve mismatches considerably below the Latin hypercube or square grid mismatches for the same numbers of training points, suggesting that the greedy algorithm is the best simulation placement strategy when the training grid cannot be tuned with trial and error.  

From a theoretical standpoint, it is not surprising that the greedy grid tends to be more accurate than these other ``pseudo-uniform'' sampling techniques.  To see this, consider the conditional covariance $K_{\rm cond}=K(X_*,X_*) - K(X_*,X)K(X,X)^{-1}K(X,X_*)$ from Equation \ref{eq:conditionalprob}. Note that as the elements of $K(X_*,X)$ get small, the second term diminishes, making the whole expression approach the prior covariance $K(X_*,X_*)$.  Assuming the training set is relatively uniform over the input space, points on the edges of the space have fewer nearby training points and hence result in smaller elements of $K(X_*,X)$.  In the center of the space, there are many nearby training points, so $K(X_*,X)$ tends to have larger elements, which decreases the elements of the conditional covariance.  In effect, the \GPR{} model has higher uncertainties near the boundaries when the training points are uniformly spread out.  Neither LH sampling nor square grids take into account that the \GPR{} uncertainties are highest near the edges.  On the other hand, the greedy algorithm accounts for the \GPR{} uncertainties and preferentially puts training points near the boundary, as can be seen in Figures \ref{fig:mismatch2D_iter} and \ref{fig:mismatch2D_est}.  

One counter-argument to using $O_k$ to measure the error is that there could be sharp features in the coefficient functions which are not sufficiently sampled by the training points and hence are poorly interpolated.  This could indeed be true in some cases, but it is an issue that applies to any interpolation scheme.  A benefit of using \GPR{} is that if sharp features exist, some of which are sampled by the training points, the hyperparameter optimization will select shorter length scales and larger covariances and hence increase the overall coefficient uncertainty across the space.  Additionally, the \GPR{} conditional distributions are Gaussian, meaning that large excursions from the mean are not ruled out --- they are just less likely.  

\section{Discussion}
\label{sec:discussion}
\begin{figure}[t]
	\centering
	\includegraphics[scale=.6]{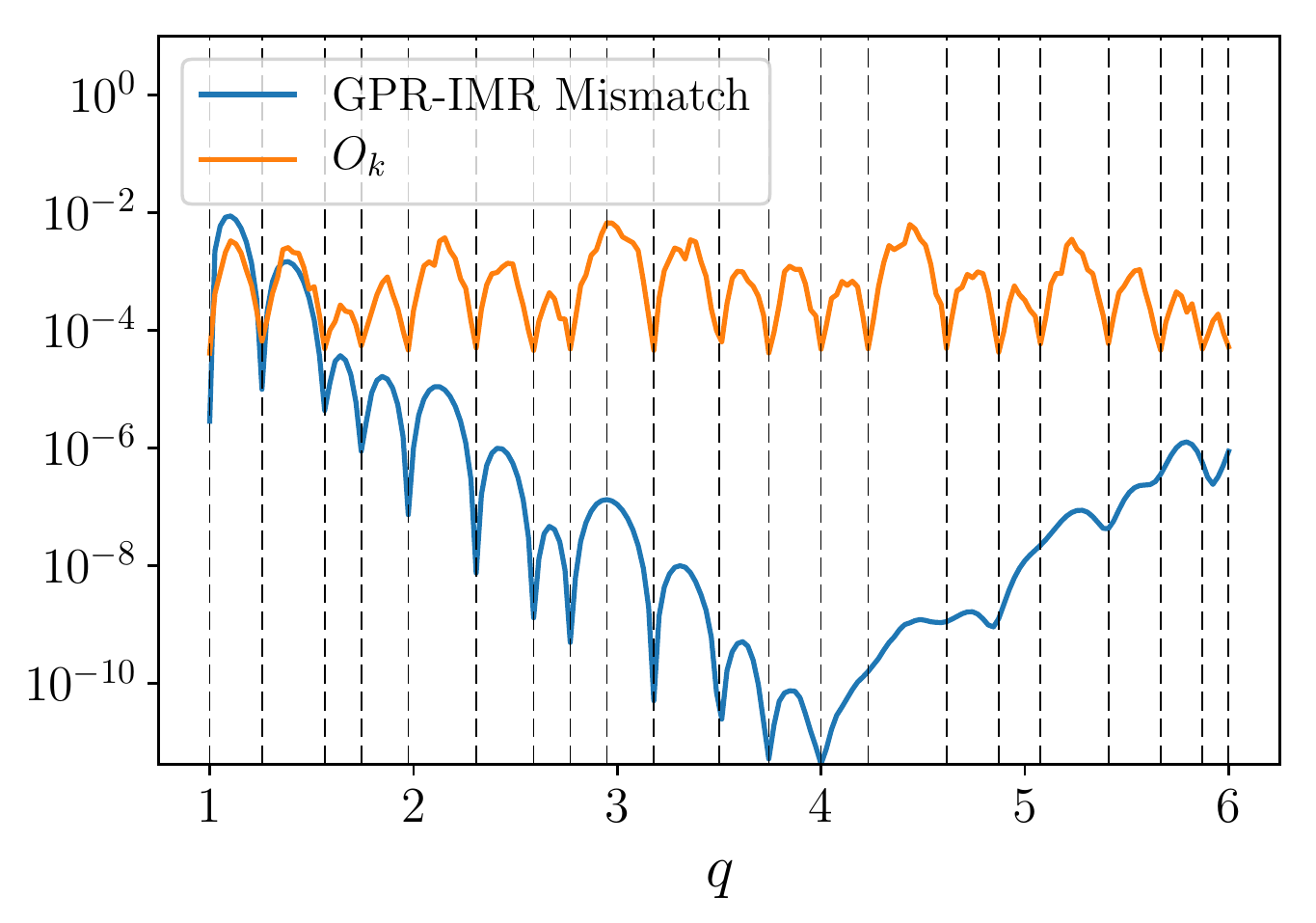}
	\caption{Mismatch between the \GPR{} mean using a Matern 5/2 kernel and IMRPhenomD (blue), and the mismatch estimated using $O_k$ based on 20 \GPR{} samples at each interpolation point (orange).}
	\label{fig:mismatch_iter_17}
\end{figure}

\begin{figure}[t]
	\centering
	\includegraphics[scale=.6]{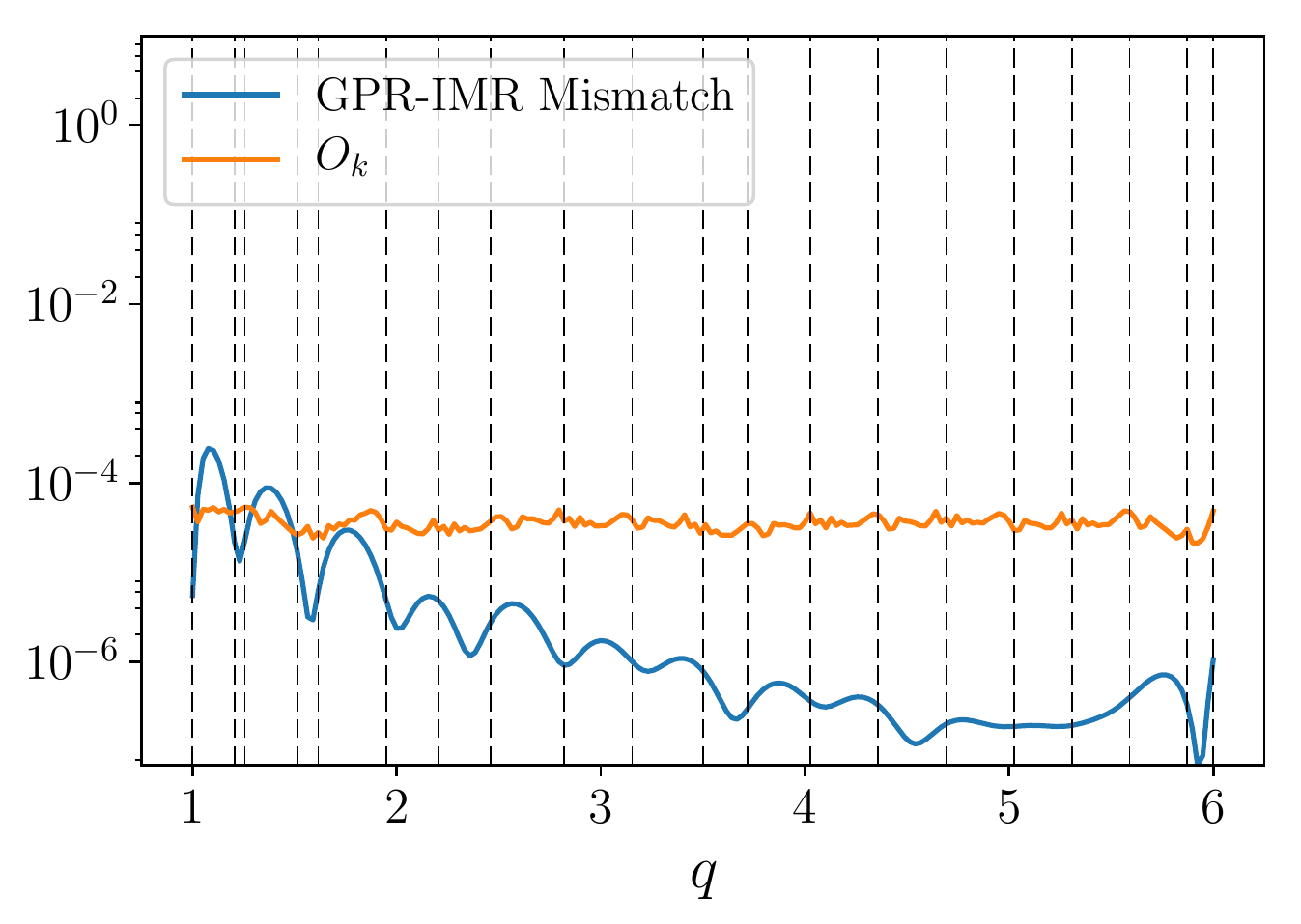}
	\caption{Mismatch between the \GPR{} mean using a squared-exponential kernel and IMRPhenomD (blue), and the mismatch estimated using $O_k$ based on 20 \GPR{} samples at each interpolation point (orange).}
	\label{fig:mismatch_iter_17_SE}
\end{figure}
In \S\ref{sec:results} we presented three examples of \GPR{}-based \ROM{} models trained on a subset of simulated waveforms. These example models produced accurate mean waveforms and quantified uncertainties across the parameter spaces of interest.  \S\ref{sec:optimize} showed that further improvements in speed and accuracy can be made to \GPR{}-based models through use of the greedy algorithm.  Although we have made specific choices in our implementation, it is to be emphasized that our method is completely general. For example, different \ROM{} or surrogate models could be used rather than the {\small SVD}-based \ROM{} we employ, hyperparameters could instead be treated as nuisance parameters and marginalized over, the \GP{} kernel could be changed, more sophisticated coefficient regularizations could be applied, and the greedy algorithm could be modified to incorporate other constraints.  We discuss a few of these possibilities here.
First, let us consider using the greedy algorithm with $O_k$ as the metric for placing new simulations.  It is worth emphasizing that greedy training point placement with $O_k$ does not strictly result in the smallest possible training set for a desired error level.  Rather, the greedy algorithm attempts to flatten the error across the parameter space by adding training points where the error is estimated to be highest. A principal limitation to just using $O_k$ to guide simulations is that it does not encode other constraints or priorities.  Two possible modifications to our greedy strategy include weighting $O_k$ by the expected simulation cost at certain parameter values and weighting by a prior on the source population parameters.  Future work will investigate these possibilities.
Next we discuss the kernel functions.  The kernel functional form encodes our assumptions about the smoothness and fluctuations of a \GP. Another common kernel choice, other than the squared-exponential kernel adopted above, is the Matern 5/2 covariance:
\beq
\mathbf{k}_{\nu=5/2}(r) = \sigma^2 \left(1+\frac{\sqrt{5}r}{l}+\frac{5r^2}{3l^2}\right)\exp\left(-\frac{\sqrt{5}r}{l}\right),
\eeq
The squared-exponential covariance constrains the \GP{} to be infinitely mean-square differentiable, imposing a strong smoothness condition on the interpolations. The Matern 5/2 kernel is less restrictive in that it only demands the process be twice mean-square differentiable. To illustrate the effects of different kernel choices, we compare with the squared-exponential covariance. For both the squared-exponential and the Matern 5/2 kernels, we apply Algorithm \ref{alg:iterative} to build a training set to interpolate waveforms for $q\in[1,6]$ as in \S\ref{sec:GPRq}.  In each case, we begin with three training waveforms to seed the algorithm: one waveform on each end of the space and one in the center.  At each iteration, $O_k$ is calculated on the fine interpolation grid from \S\ref{sec:GPRq} and a new point is placed at the point of highest $O_k$.  

Figure \ref{fig:mismatch_iter_17} shows the real \GPR-IMR mismatch and the mismatch estimated from $O_k$ using the Matern 5/2 kernel, and Figure \ref{fig:mismatch_iter_17_SE} shows the same for a squared-exponential kernel.  In both cases, the maximum $O_k$ over the space bounds the real mismatch, except near $q=1$ where the coefficient functions vary faster than elsewhere in the space (see Figures \ref{fig:amp_coeffs} and \ref{fig:phase_coeffs}).  The squared-exponential kernel is able to keep the error relatively constant over the space.  Additionally, the maximum mismatch using the squared exponential is about two orders of magnitude lower than when using the Matern 5/2 kernel.  On the other hand, use of the Matern kernel results in significantly lower error on the interior of the space. Further study of kernel effects will be required as new portions of parameter space are explored with the \GPR{} model.  In particular, higher dimensional \GPR{} models may require more complex kernels, which can be constructed by summing or multiplying pre-existing kernel functions.  Additionally, kernels with compact support should be considered, as they can allow faster \GP{} evaluation and enhanced computational stability.

We now shift our focus to the execution time of generating \GPR{} waveforms.  The examples shown here were designed to run in less than one day on one computing node with 16 cores, but future work would make use of more cores, allowing, for example, a larger ROM basis, or interpolation and $O_k$ calculation on a finer grid. To illustrate the scaling of required time and resources with the number of \GP{} training points, we perform a \GPR{} on one coefficient in the $q$-$\chi$ space for different numbers of training values and evaluation points.  The numbers of training and evaluation points determine the sizes of the matrices that must be multiplied in Equation \ref{eq:conditionalprob} and hence the execution time.  It is worth noting that $K(X,X)^{-1}$ or $K(X,X)^{-1}{\mathbf f}$ with optimized hyperparamters can be pre-calculated, allowing the most computationally intensive step in the \GPR-building to be done just once ahead of time.  We assume here that the optimization and matrix inversion steps have already been done and simply look at the evaluation time of a \GP{}. 

\begin{figure}
	\centering
	\includegraphics[scale=.5]{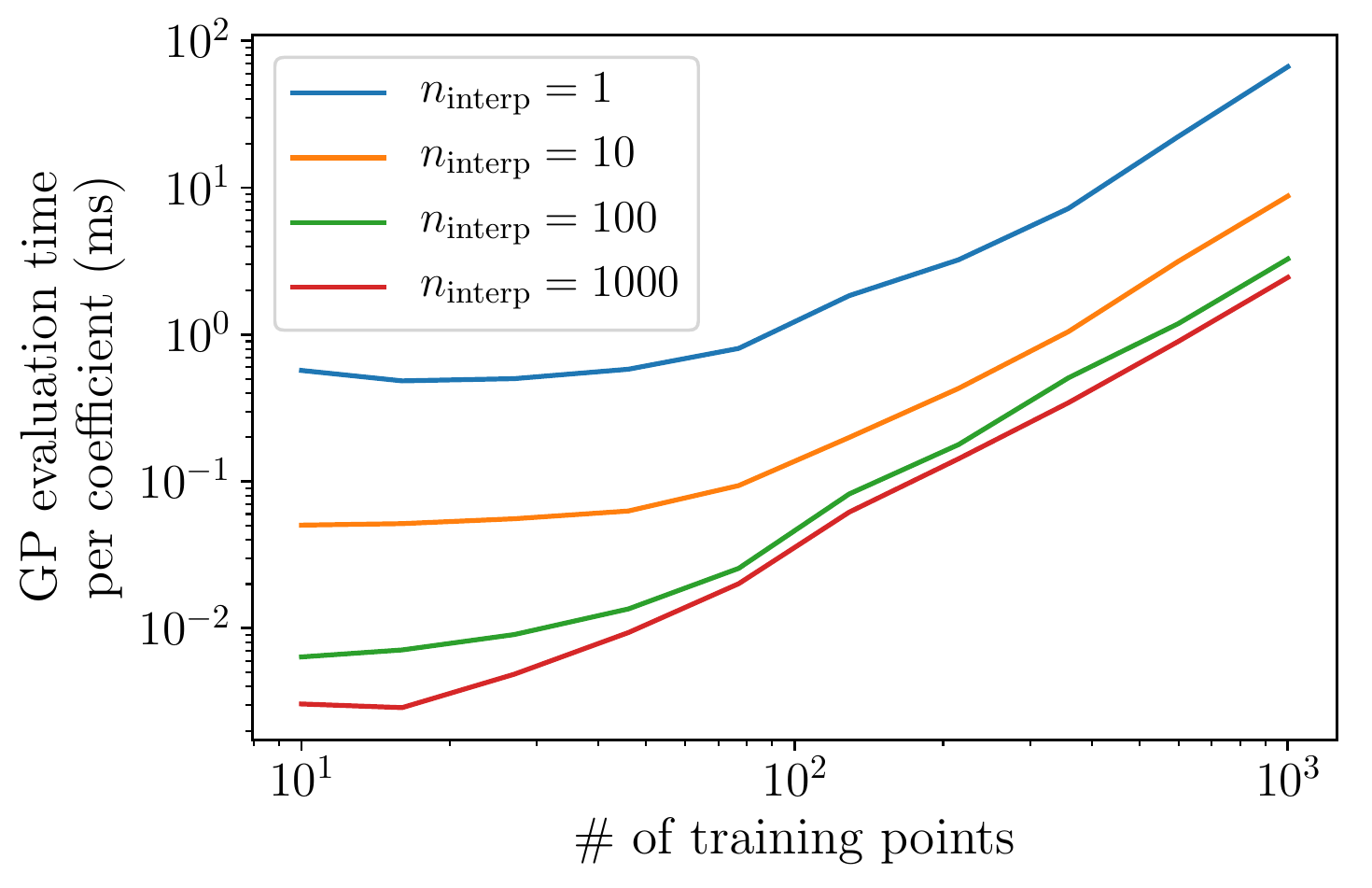}
	\caption{Time to evaluate (mean and variance) one coefficient per interpolation point in two dimensions as a function of the number of training points.  Times shows are from \GP{} evaluations on a 2.6GHz Intel E5-2670 CPU.}
	\label{fig:timing}
\end{figure}

The timing results are shown in Figure \ref{fig:timing}, which plots the conditional \GP{} mean and covariance evaluation time per coefficient per interpolated point at $n_{\rm interp}$ points as a function of the number of training points. That is, we evaluate the \GP{} mean and covariance at $n_{\rm interp}$ points and divide the total time by $n_{\rm interp}$ to show the time per interpolated coefficient value. In principle, the total evaluation time should scale directly with the number of interpolated points, but Figure \ref{fig:timing} indicates that interpolating more points at once gives an overall speedup.  This is due to the overhead in constructing the training-test covariance matrices in the \texttt{scikit-learn} implementation of the \GPR{}.  This overhead is further evidenced by the fact that the $n_{\rm interp} = 100$ and $n_{\rm interp} = 1000$ curves in Figure \ref{fig:timing} converge. Future work will consider alternate \GPR{} implementations to mitigate such overhead, since typically only single waveform evaluations are required.  

Note that to build a full \GPR{} waveform, a \GP{} must be evaluated for each of the $\sim 100$ phase and amplitude coefficients.  In the case that the overhead cannot be bypassed ($n_{\rm interp} = 1$, blue curve), it would take $\sim 1$ minute per waveform evaluation, assuming 100 coefficients and 1000 training points.  If the overhead can be entirely removed ($n_{\rm interp} = 1000$, red curve), each waveform evaluation would instead take $\sim 200$ms. This can be compared to the evaluation time for the spline-based \ROM{} in P14 of $\sim 1$ ms depending on the system's total mass (see Figure 1 of P14). Further speedups to our model could be achieved by lowering the number of \ROM{} coefficients, decreasing the number of training points, or evaluating the coefficients in parallel. To decrease the number of training points, domain decomposition could be used as suggested in \S\ref{sec:2DGPR}. If domains of $\sim 100$ training points were used, the \GP{} evaluation times would fall by over an order of magnitude (evaluation time goes as $n_{\rm train}^2$).  Another possibility is to apply the subset of regressors method, which effectively considers a only a subset of rows of $K(X,X)^{-1}\mathbf{f}$ in Equation \ref{eq:conditionalprob}.  This scheme reduces the size of the matrices multiplied when calculating the \GP{} conditional mean and covariance (see \S8.3.1 in \citep{Rasmussen}). 

As mentioned before, the most expensive step of building a \GPR{} model is in training, where the hyperparameters must be optimized and $K(X,X)$ must be inverted. The matrix inversion computation time scales as $O(n_{\rm train}^3)$, since $K(X,X)$ is an $n_{\rm train} \times n_{\rm train}$ matrix (see e.g. \citep{Rasmussen}).  Given the timescale for generating an \NR{} waveform, we would not expect more than $O(10,000)$ \NR{} waveforms in the near future, so the matrix inversion step should not be prohibitively expensive, especially since the kernel for each coefficient can be handled in parallel, and a $O(1000)$ training set can be optimized on a personal computer in minutes and only needs to be done once.  Also, as mentioned earlier, smaller domains with fewer training waveforms can be handled with separate \GPR{} models or a subset of regressors can be used, mitigating the need for the inversion of very large covariance matrices.

\medskip
\section{Conclusions}
\label{sec:conclusion}
We have demonstrated that Gaussian process regression can be used to construct reduced-order-model waveforms with uncertainties using only a few existing simulations, and that these uncertainties can guide the choice of future simulations.  The overall motivation for such \GPR{} models is that \GPR{} uncertainties can be propagated to the parameter estimation of compact binary coalescences in order to remove bias in estimates due to systematic waveform errors. Figures \ref{fig:amp_sample} and \ref{fig:phase_sample} show example amplitude and phase functions with uncertainties from the \GPR{} model. 

This work has also shown that \GPR{} can model waveforms accurately over a parameter space of interest.  Figures \ref{fig:mismatch2D} and \ref{fig:mismatch2D_iter} show that with a sufficient number of training simulations, the error level of the \GP{} model can be reduced to levels adequate for parameter estimation with \LIGO{} data, especially if Algorithm \ref{alg:iterative} is used to construct the training set.  Such greedy algorithms will be a particularly useful tool for efficiently choosing the parameters of new simulations in the nominal 7-d parameter space of interest to \LIGO{}.  Since the methods presented here are general, they could in principle be applied to other scenarios such as eccentric-orbit Laser Interferometer Space Antenna sources or neutron-star binaries with tidal deformability parameters. 

Another finding of this work is that the error level of the \GPR{} model can be estimated from the \GPR{} itself rather than through cross-validation.   Figure \ref{fig:error_iter}, which compares the maximum true \GPR{} error to the maximum estimated error, demonstrates that the \GPR{} uncertainties can alone be used to estimate the maximum error level of the \GPR{} model.  This allows one to know when a \GPR{} model has reached a desired error level and does not require further simulations in the parameter region of interest.

Finally, we describe future directions for this work. In the immediate future, \GPR{} models will be applied to three- or higher-dimensional parameter spaces to test the robustness of these models as the complexity grows. In particular, we will see if the \ROM{} coefficient functions can generally be treated as being uncorrelated.  Additionally, a wider range of kernel functions will be explored than what has been presented here. In the longer term, \GPR{} training sets will be built directly from \NR{} simulations, rather than utilizing a stand-in approximant.  Apart from doing \PE{} studies with an \NR-based \GPR{} model, the model could also be used to validate or study families of approximants.  In sum, \GPR{} models present an exciting frontier for \NR-simulation-driven models of \GW{} waveforms.

\section{Acknowledgements}
DEH acknowledges valuable discussions with Salman Habib, Katrin Heitmann, David Higdon, and Michael Stein.  ZD would like to thank Richard Chen for consultation on \GPR{}. ZD is supported by NSF Graduate Research Fellowship grant DGE-1144082. ZD, BF, and DEH were partially supported by NSF CAREER grant PHY-1151836 and NSF grant PHYS-1708081. They were also supported in part by the Kavli Institute for Cosmological Physics at the University of Chicago through NSF grant PHY-1125897 and an endowment from the Kavli Foundation. This work was completed in part with resources provided by the University of Chicago Research Computing Center.


\bibliography{GP}
\end{document}